\documentclass[11pt,twoside]{article}

\usepackage{mathtools} 
\usepackage{mathrsfs} 
\usepackage[symbol]{footmisc}

\newcommand{\h}{\mathfrak{h}}
\newcommand{\hn}{\mathfrak{h}_{\mathrm{ph}}}

\newcommand{\DD}{\mathcal{D}}
\newcommand{\WW}{\mathcal{W}}
\newcommand{\FF}{\mathcal{F}}
\newcommand{\HH}{\mathscr H}
\newcommand{\HHn}{\mathscr H_{\mathrm{nel}}}
\newcommand{\HHel}{\mathscr H_{\mathrm{el}}}
\newcommand{\HHex}{\mathscr H_{\mathrm{ex}}}
\newcommand{\tHH}{\tilde{\mathscr H}}
\newcommand{\LL}{\mathscr L}
\newcommand{\tH}{\tilde{H}}
\newcommand{\Hel}{H_{\mathrm{el}}}
\newcommand{\Hn}{H_{\mathrm{nel}}}

\newcommand{\Hex}{H_{\mathrm ex}}
\newcommand{\Hext}{\tilde{H}_{\mathrm ex}}
\newcommand{\tR}{\tilde{R}}
\newcommand{\hh}{\mathfrak{h}}

\newcommand{\chizero}{\rchi_{\{0\}}(H_-)}

\newcommand{\uGamma}{\breve{\Gamma}}
\newcommand{\udGamma}{\mathrm{d}\breve{\Gamma}}

\newcommand{\ran}{\mathrm{Ran}}

\usepackage{amsmath,amssymb,amsthm}
\usepackage{pstricks,pst-node,pst-coil,pst-plot,pstricks-add}
\usepackage{geometry,epsfig}
\usepackage{bbm}
\usepackage{cite}

\usepackage{cite}
\usepackage{paralist}
\usepackage{cleveref}
\usepackage{enumitem}
\usepackage{emptypage}

\DeclareGraphicsExtensions{.pdf}

\usepackage{pgfplots}
\pgfplotsset{compat=1.18}
\usepackage{amsmath}

\newtheorem{lemma}{Lemma}[section]
\newtheorem{theorem}[lemma]{Theorem}

\newtheorem{proposition}[lemma]{Proposition}

\newcommand{\C}{\mathbb{C}} 
\newcommand{\R}{\mathbb{R}} 
\newcommand{\N}{\mathbb{N}} 

\DeclareRobustCommand{\rchi}{{\mathpalette\irchi\relax}}
\newcommand{\rchinull}{\rchi_{\{0\}}}
\newcommand{\irchi}[2]{\raisebox{\depth}{$#1\chi$}} 

\newcommand{\eps}{\varepsilon}
\newcommand{\ph}{\varphi}
\newcommand{\supp}{\operatorname{supp}}
\newcommand{\const}{\operatorname{const}}

\newcommand{\sprod}[2]{\langle #1, #2\rangle}
\newcommand{\expect}[1]{\langle #1\rangle}

\title{\textbf{On Rayleigh scattering in the massless\\ Nelson model}}

\author{M.~Griesemer\footnote{marcel.griesemer@mathematik.uni-stuttgart.de}\,\ and V.~Ku{\ss}maul\footnote{valentin.kussmaul@mathematik.uni-stuttgart.de}\\  
\small Fachbereich Mathematik, Universit\"at Stuttgart, D-70569 Stuttgart, Germany}  
\date{}

\begin{document}
\maketitle

\vspace{-10mm}
\begin{center}
\emph{Dedicated to Israel Michael Sigal on the occasion of his eightieth birthday.}
\end{center}

\begin{abstract}
Asymptotic completeness of Rayleigh scattering in models of atoms and molecules of non-relativistic QED is expected, but for a proof we still lack sufficient control on the number of emitted soft photons. So far, this obstacle has only been overcome for the spin-boson model. In a general class of models asymptotic completeness holds provided the expectation value of the photon number $N$ remains bounded uniformly in time. This has been established by Faupin and Sigal. 
We review and simplify their work, and, more importantly, we replace the bound on $N$ by a weaker assumption on the distribution of $N$ that is both necessary and sufficient for asymptotic completeness.
\end{abstract}

\section{Introduction} \label{introduction}

Atoms and molecules in excited states with energy below the ionization threshold relax to the ground state by emission of the excess energy in terms of photons. In mathematical models such a phenomenon is expected to occur under fairly general circumstances. Existence of a ground state, instability of excited states, and a certain decay of correlations seem to be sufficient \cite{DGK2015}. Yet the only proofs known so far concern simplified models such as the spin-boson model, the Pauli-Fierz model with an infrared cutoff or massive bosons, explicitly solvable models and perturbations thereof \cite{Arai1983,Spohn1997,DG1999,FGSch2002,DGK2015}. In a general setting, the lack of sufficient control on the number of emitted photons is the obstacle. Taking such a control for granted, asymptotic completeness has been proven in a remarkable paper by Faupin and Sigal \cite{FauSig2014}. The purpose of the present work is twofold: first, we show that the strategy of \cite{FauSig2014} can be implemented with much less effort, by working in an expanded Fock space containing additional fake bosons of negative energy, an idea due to Jak\v{s}i\'c and Pillet. Second, we derive a new propagation estimate, which allows us to work with a weaker a priori assumption on the number of emitted photons that is both necessary and sufficient for asymptotic completeness.

In this paper we consider a one-electron atom described by a regularized Nelson Hamiltonian with massless bosons. Our methods equally apply to (generalized) spin-boson models and, with more work, to many-electron Pauli-Fierz Hamiltonians. The Hilbert space of the system is the tensor product  $\HHn = \HHel \otimes \FF(\hn)$ of the one-electron space $\HHel = L^2(\R^3, dx)$ and the symmetric Fock space $ \FF(\hn)$ over the one-boson space $\hn = L^2(\R^3, dk)$. The Hamiltonian has the form
\begin{align} \label{Nelson-Hamiltonian}
  \Hn = \Hel \otimes 1 + 1\otimes H_\omega + g\phi(w_x).
\end{align}
The first term, $\Hel = -\Delta + V$, is a Schr\"odinger operator with potential $V : \R^3 \to \R$ satisfying $V_+ \in L_\mathrm{loc}^2(\R^3)$ and $V_- \in L^2(\R^3) + L^\infty(\R^3)$.  These conditions on $V$ are sufficient to define a self-adjoint operator $\Hel$ via a semi-bounded closed quadratic form.
We assume that $e_0 = \inf \sigma(\Hel)$ is a simple eigenvalue below the essential spectrum of $\Hel$, which is the case for the typical potentials we have in mind, such as Coulomb potentials or confining potentials.

The second term in \eqref{Nelson-Hamiltonian} is the operator of the field energy $H_\omega = d\Gamma(\omega)$, where $\omega$ denotes multiplication with $\omega(k ) = |k|$ in the one-boson space. The last term accounts for the particle-field interaction. The parameter $g>0$ denotes a coupling constant and 
$$\phi(w_x) = a^*(w_x) + a(w_x),$$
with $ a^*(w_x)$ and $a(w_x)$ denoting the usual creation and annihilation operators in Fock space $\FF(\hn)$. We assume that $w_x(k) = e^{-i\langle k,x\rangle} w(|k|)$ with 
$x\in \R^3$ the position of the electron and $w(\omega) = \omega^\mu \zeta(\omega)$. The Schwartz function $\zeta \in \mathcal{S}(\R)$ describes an ultraviolet cutoff and should be thought of as a constant near $\omega = 0$. For $\mu > -1$ the Hamiltonian $\Hn$ is self-adjoint with domain $D(\Hn) = D(\Hel \otimes 1 + 1 \otimes H_\omega)$ and bounded from below with spectrum $\sigma(\Hn) = [E, \infty)$. Existence of a ground state $\psi_{gs} \in \HHn$, $\Hn\psi_{gs} = E\psi_{gs}$, requires that  $\mu > -1/2$; our main result assumes $\mu>1/2$. Since $g$ needs to be sufficiently small in our main result, we may assume without loss of generality that the ground state is unique \cite{BFS1998, DG2004, G2000, LMS2002}. 

To avoid the possibility of ionization, the energy distribution of the initial state must be bounded above by the so-called \textit{ionization threshold} 
\begin{align*}  
	\Sigma = \lim_{R \to \infty} \inf_{\psi \in D_R} \sprod{\psi}{\Hn\psi},
\end{align*} 
where $D_R$ consists of all normalized states $\psi \in D(\Hn)$ satisfying $\rchi(|x| < R) \psi = 0$. Our assumptions on $\Hel$ guarantee that
$\Sigma - E > 0$ for $g$ small enough. If we choose $\eps>0$ with $\eps^2 < \Sigma - E$ then $e^{\eps|x|} f(\Hn)$ is bounded for all $f \in C_0^\infty(-\infty, \Sigma)$ \cite{G2004}. This confirms the picture of a localized electron.

\textit{Asymptotic completeness} in some interval $\Delta =[E,\lambda)$, $\lambda<\Sigma$, holds if every vector $\psi\in \ran \, \rchi_\Delta(\Hn)$ 
is an \emph{(outgoing) scattering state} of $\Hn$ in the following sense: for all $\eps>0$ there exist $\eta\in \FF(\hn)$ with $\eta^{(n)}=0$ for almost all $n\in \N$, and $T>0$ such that 
\begin{equation}\label{acp}
    \big\| e^{-i\Hn t}\psi - I\left(e^{-iEt} \psi_{gs} \otimes e^{-iH_\omega t}\eta\right) \big\| <\eps\qquad \text{all}\ t>T.
\end{equation}
Here, the \textit{scattering identification} $I$ is an operator that merges bosons from the second factor with the first.  Somewhat formally, this can be defined by
$$
       I\left(\psi_{gs} \otimes \eta\right) = \sum_{n\ge 0} \frac{1}{\sqrt{n!}} \int \eta^{(n)}(k_1,\ldots,k_n) a^{*}(k_1)\cdots a^{*}(k_n)\psi_{gs}\, dk_1\ldots dk_n.
$$
Asymptotic completeness as described above requires the instability of excited states, which is usually expressed in terms of the \emph{Fermi golden rule} (FGR) condition. For our purpose it is more efficient to work with the following consequence of Mourre-theory, which, depending on $\lambda$, implicitly requires an FGR condition to hold:
\begin{description}
\item[(v)] \label{v} For all $f\in C_0^{\infty}(E,\lambda)$, $s<1/2$, $g>0$ small enough, and $\psi\in \HHn$,
\begin{align*}
      \| \rchi(N = 0) e^{-i\Hn t} f(\Hn)\psi \| = O(t^{-s}) \|\expect{A}\psi\| \quad (t \to \infty),
\end{align*}
where $A = d\Gamma(a)$, $a = (k \cdot i \nabla_k + i \nabla_k \cdot k)/2$ denotes the dilation generator, and $\expect{A}=(1+A^2)^{1/2}$. The operator $N = d\Gamma(1)$ is the number operator and $\rchi(N = 0)$ the vacuum projector. 
\end{description}
 In \cite{FauSig2014}, Lemma 4.3, property (v) is established for a class of functions $f\in C_0^{\infty}(E,\Sigma)$ whose support can be covered with finitely many intervals for which a Mourre estimate holds. The required Mourre estimates are established in \cite{FGS2008,BFSS1999}, where FGR is assumed 
on the excited states of the non-interacting system. In the vicinity of the ground state energy $E$ there are no eigenvalues of such excited states and hence no FGR assumption is needed \cite{FGS2008}.

With these preparations we can now state our main result:

\begin{theorem} \label{N-implies-AC}
Suppose $\mu>1/2$, $\lambda<\Sigma$, $g>0$ is small enough and \textnormal{(v)} 
holds\footnote[1]{Notice that the assumption $\langle g\rangle \ll 1$ in \cite{FauSig2014}, Theorem 1.1,  requires $\mu>1/2$ as well.}. 
Let $\Delta = [E, \lambda)$. Then a state $\psi\in \ran \, \rchi_{\Delta}(\Hn)$ is a scattering state of $\Hn$ in the sense of \eqref{acp} if and only if
\begin{equation}\label{N-bound}
      \|\rchi(N\geq m)e^{-i\Hn t}\psi\| \to 0\qquad (m,t\to\infty).
\end{equation}
\end{theorem}

\noindent
\emph{Remarks.}
\begin{enumerate}
\item For fixed $t$ it is clear that $ \|\rchi(N\geq m)e^{-i\Hn t}\psi\| \to 0$ as $m\to\infty$. The point of \eqref{N-bound} is the uniformity in large $t$.
\item The set of all states $\psi\in\HHn$ with property \eqref{N-bound} is a closed, non-empty subspace, which is invariant under $\Hn$. Therefore, by \Cref{N-implies-AC}, to prove asymptotic completeness in $\Delta$ it suffices to verify that \eqref{N-bound} holds for all $\psi$ from \emph{some} dense subset of  $\ran \, \rchi_{\Delta}(\Hn)$.
\item For the spin-boson model, it is known that $\sup_{t>0}\sprod{\psi_t}{N\psi_t}<\infty$ for $\psi$ from a suitable dense subspace of the Hilbert space \cite{DK2013}. In view of the remark above and the Chebyshev inequality,  $\|\rchi(N\geq m)\psi_t\|^2 \leq \sprod{\psi_t}{N\psi_t}/m$, we conclude that \eqref{N-bound} holds for all $\psi$ in the spin-boson model and hence, by \Cref{N-implies-AC}, asymptotic completeness follows. This was previously shown in \cite{DGK2015}, which also builds upon \cite{DK2013}. 
\item For the Nelson model with an infrared cutoff, we have $\sup_{t>0}\sprod{\psi_t}{N\psi_t}<\infty$ for $\psi\in D(N^{1/2})\cap D(|\Hn|^{1/2})$. Hence, by the remarks above, condition \eqref{N-bound} is satisfied and asymptotic completeness follows, see also \cite{FGSch2002}.  Without infrared cutoff we only know that $\| \rchi(N \geq t^\nu) \psi_t \|\to 0$ as $t \to \infty$ if $\nu > 1/(2 + \mu)$, see \Cref{Gerard bound} below, which does not seem quite sufficient.
\end{enumerate}

The present paper is inspired by \cite{FauSig2014, Gerard2002}. It builds and expands upon ideas and methods from these papers.
In \cite{FauSig2014} asymptotic completeness is derived from the assumption that $\sup_{t>0}\sprod{\psi_t}{N\psi_t} \le C\sprod{\psi}{(N+1)\psi}$ for $\psi \in f(\Hn)D(N^{1/2})$ with $f\in C_0^{\infty}((E,\Sigma))$ and $C$ independent of $\psi$.
Alternatively, a similar bound on $\sprod{\psi_t}{d\Gamma(\omega^{-1})\psi_t}$ for initial states from a fairly general dense subspace of
$\ran E_{(-\infty,\Sigma)}(H)$ is shown to be sufficient. 
The paper \cite{Gerard2002} contains interesting partial results towards asymptotic completeness for confined Nelson models. These results involve spaces  $\HH_{c}^{+}$ with finitely many bosons in $\{r\ge ct\}$ as $t\to\infty$, with $r$ the photon position and $c<1$. If one assumes that $\HH_{c}^{+}$ agrees with the entire Hilbert space, which is a natural assumption similar to \eqref{N-bound}, then a weak form of asymptotic completeness, where asymptotic vacua play the role of ground states, can be inferred from Theorem 12.3 (iv) in \cite{Gerard2002}.

The main elements of the proof of \Cref{N-implies-AC} are a suitable Deift-Simon wave operator $W$, \Cref{thm-W}, and a minimal escape property, 
\Cref{min-escape}, which are derived from condition \eqref{N-bound} and Hypothesis (v), respectively. The proof of \Cref{N-implies-AC} based on these elements is patterned after the proof of AC in \cite{FauSig2014}. In contrast to \cite{FauSig2014} we work in the extended Fock space, $\FF(L^2(\R\times S^2))$, which is isomorphic to $\FF(L^2(\R^3))\otimes \FF(L^2(\R^3))$ via the Jak\v{s}i\'c-Pillet glueing trick. The bosons from the second Fock space are fake bosons with negative energy. The advantage of this enlarged system is that the field energy in $\FF(L^2(\R\times S^2))$ is the operator $d\Gamma(s)$, where $s$ denotes multiplication with the first argument of a function in $L^2(\R\times S^2)$. Since we take $r=i\partial/\partial s$ for the position operator, it becomes very easy to control commutators of $s$ with localization functions $j(r)$. The localization of bosons in the original Fock space $ \FF(L^2(\R^3))$ is made difficult by the fact that $\omega(k)$ is $\sqrt{-\Delta}$ in position space. In \cite{FauSig2014} a lot of work is devoted to this problem.

Our main progress, on a technical level, is a new propagation estimate, \Cref{main propagation estimate}, which allows us to replace the uniform bound on $\sprod{\psi_t}{N\psi_t}$ by the weaker assumption \eqref{N-bound}. This new propagation estimate is inspired, in part, by results from \cite{Gerard2002}. A further improvement compared to  \cite{FauSig2014} is that \Cref{N-implies-AC} asserts a state-wise connection between hypothesis and result, that is, \eqref{N-bound} for a given $\psi\in \ran \, \rchi_{\Delta}(\Hn)$ is equivalent to \eqref{acp} for that $\psi$.  

This paper is organized as follows. In \Cref{sec:necessity} we prove the necessity of condition \eqref{N-bound}. In \Cref{sec:expanded} we introduce the expanded system, and we rewrite Hypothesis (v) and \Cref{N-implies-AC} in terms of operators of the expanded system, see \Cref{vac-decay} and \Cref{thm:AC}. The rest of the paper is devoted to the proof of \Cref{thm:AC}. \Cref{sec:DS} establishes existence of the Deift-Simon operator $W$, see \Cref{thm-W}, with the help of the propagation estimates \Cref{dn prop} and \Cref{main propagation estimate}. \Cref{sec:min-escape} is devoted to the minimal escape property \Cref{min-escape}, which follows from Hypothesis (v). Finally, in \Cref{main thm proof}, we combine all ingredients to prove \Cref{thm:AC}. There are various appendices collecting background information on second quantization and auxiliary results.


\section{Decay of the $N$-distribution is necessary}
\label{sec:necessity}

In this section we show that condition \eqref{N-bound} on the distribution of $N$ is necessary for a state to be a scattering state of $\Hn$. This is the easy part of \Cref{N-implies-AC}. 

Let $\Omega \in \FF(\hn)$ be the vacuum state and let $\FF_\mathrm{fin}(\hn)\subset \FF(\hn)$ denote the finite particle subspace.
We define the \textit{scattering identification} $I$ as the closure of the operator $I : \FF_\mathrm{fin}(\hn) \otimes \FF_\mathrm{fin}(\hn) \rightarrow \FF(\hn)$ characterized by 
\begin{align}
	a^*(h_1) ... a^*(h_k) \Omega \otimes a^*(g_1) ... a^*(g_\ell) \Omega \mapsto  a^*(g_1) ... a^*(g_\ell) a^*(h_1) ... a^*(h_k) \Omega. \label{Creation def for I}
\end{align} 
For an alternative definition of $I$ see \Cref{sec:Fock}. The closed operator $I: D(I) \subset \FF(\hn) \otimes \FF(\hn) \rightarrow \FF(\hn)$ is unbounded, but from \eqref{Creation def for I} it is easy to see that for every $n \in \N_0$ 
\begin{align}  \label{I relative bound}
	I \left((N + 1)^{-n/2}\otimes \rchi(N \leq n) \right) 
\end{align} 
is a bounded operator. Since $\psi_{gs}$ belongs to the domain of any power of the number operator $N$, see \cite{BHLMS2001}, it follows that $I (\psi_{gs} \otimes \eta)$ is well-defined for all $\eta \in \FF_\mathrm{fin}(\hn)$. At the expense of restrictions on the class of admissible $\eta$, we could work with powers of the field energy $H_\omega$, rather than powers of $N$, and avoid the use of  \cite{BHLMS2001}.

\begin{lemma} \label{AC implies N}  \, \\ 
\vspace{-6mm}
\begin{enumerate}
\item[(i)] Let $\eta \in \FF_\mathrm{fin}(\hn)$. Then
\begin{align*}
	\sup_{t \in \R}\| \rchi(N \geq m) I (e^{-i E t} \psi_{gs} \otimes e^{-i H_\omega t}\eta) \| \longrightarrow 0 \quad (m \to \infty).
\end{align*}

\item[(ii)] If $\psi \in \HHn$ is a scattering state in the sense of \eqref{acp} then $\rchi(N \geq m) e^{-i \Hn t}\psi \to 0$ as $m, t \to \infty$. 
\end{enumerate}
\end{lemma}
\begin{proof}
(i) From \eqref{Creation def for I} it follows that for all $m, n \in \N_0$
\begin{align*}  
	\rchi(N \geq m) I \big( 1 \otimes \rchi(N \leq n) \big) = \rchi(N \geq m) I \big( \rchi(N \geq m - n) \otimes \rchi(N \leq n) \big).
\end{align*} 
Hence, if $\eta = \rchi(N \leq n) \eta$ for some $n \in \N$ then, by \eqref{I relative bound},
\begin{align*}  
	\lefteqn{\| \rchi(N \geq m) I (e^{-i E t} \psi_{gs} \otimes e^{-i H_\omega t}\eta) \|}  \\ 
	&\leq \|I\left( (N + 1)^{-n/2}\otimes \rchi(N \leq n) \right)\|  \, \| \rchi(N \geq m - n) (N + 1)^{n/2}\psi_{gs} \| \, \| \eta \| \\ 
	&\longrightarrow 0 \quad (m \to \infty).
\end{align*} 
(ii) If $\psi$ is a scattering state, then there exists $\eta \in \FF_\mathrm{fin}(\hn)$ such that, for large times, $e^{-i \Hn t} \psi$ is well approximated by $I (e^{-i E t} \psi_{gs} \otimes e^{-i H_\omega t}\eta)$. In view of $\| \rchi(N \geq m) \| \leq 1$ and (i) it follows that $\rchi(N \geq m) e^{-i \Hn t}\psi \to 0$ as $m, t \to \infty$. 
\end{proof}

\section{The expanded system} 
\label{sec:expanded}

In this section we introduce the \textit{expanded system} and we reformulate hypotheses and \Cref{N-implies-AC} in terms of objects of this system.
For the motivation of this step we refer to the introduction. We begin by defining some auxiliary operators, to be used for relating the expanded system to the original one. In the remainder of the paper, only the operators $H_+, H_-$ and $H$, see \eqref{DEFINITION H}-\eqref{DEFINITION H_-}, as well as the reformulation of \Cref{N-implies-AC} in the form of \Cref{thm:AC} are needed.

Let $H_\omega = d\Gamma(\omega)$ for short, and let
\begin{align*}
     \Hex &\coloneqq \Hn\otimes 1 - 1\otimes H_{\omega}\quad \hspace{25.5mm} \text{in } \HHex \coloneqq \HHn \otimes \FF(\hn), \\
     \Hext &\coloneqq \Hex \otimes 1 + 1\otimes (H_{\omega}\otimes 1- 1\otimes H_\omega)\quad \text{in } \tilde{\HHex} \coloneqq \HHn \otimes \FF(\hn) \otimes \FF(\hn) \otimes \FF(\hn).
\end{align*}
Let $I_\mathrm{ex}: D(I_\mathrm{ex}) \subset \tilde{\HHex} \to \HHex$ be the closure of the operator
\begin{align*}  
	I_\mathrm{ex} \big( \psi \otimes \mu \otimes \eta \otimes \nu \big) = \left(I(\psi \otimes \eta)\right) \otimes \big(I (\mu \otimes \nu) \big)
\end{align*} 
with $I$ the scattering identification \eqref{Creation def for I}. The operator $I_\mathrm{ex}$ merges bosons of positive and bosons of negative energy, respectively. 
We can now say that $\psi \in \HHn$ is a scattering state of $\Hn$ in the sense of \eqref{acp} if and only if $\psi \otimes \Omega$ is a scattering state of $\Hex$, that is, for all $\eps > 0$ there exists $\eta \in \FF_\mathrm{fin}(\hn)$ and $T > 0$ such that for all $t > T$
\begin{align} \label{AC reformulated with Hex}
           \big\| e^{-i \Hex t}(\psi \otimes \Omega) - I_\mathrm{ex} e^{-i \Hext t} (\psi_{gs} \otimes \Omega \otimes \eta \otimes \Omega )\big\| < \eps. 
\end{align} 
Indeed, the norm in \eqref{AC reformulated  with Hex} agrees with the norm in \eqref{acp}.

It is clear that the dynamics of the negative energy bosons is irrelevant. The point of our particular choice is that $\FF(\hn) \otimes \FF(\hn)$ can be mapped onto the Fock space $\FF(\h)$ over $\hh \coloneqq L^2(\R \times S^2, ds \, dS(\sigma))$, where the combined free dynamics takes a very simple form. This construction is described in the following and well known from \cite{JP1996, Gerard2002, DJ2001}. 

With $U^{*}: \FF(\hn) \otimes \FF(\hn) \to \FF(\hn \oplus \hn)$ denoting the adjoint of the canonical unitary \eqref{canonical unitary}, we have 
\begin{align} 
 	U^* \big( H_\omega \otimes 1 \big) &= d\Gamma(\omega \oplus 0) U^*, \label{U1} \\ 
	U^* \big( 1 \otimes H_\omega \big) &= d\Gamma(0 \oplus \omega) U^*, \\ 
      U^{*} \big(\phi(w_x)\otimes 1 \big) &= \phi(w_x,0) U^{*}. \label{U2}
\end{align}
We define the unitary operator $V : \hn \oplus \hn \to \hh $ by
$$
              V(f,g)(s,\sigma) := \begin{cases} sf(s\sigma) & s \geq 0, \\   sg(-s\sigma)& s < 0,\end{cases}\quad  \quad (s, \sigma) \in \R \times S^2.
$$  
Then, with $s_\pm  = \max(\pm s, 0)$, 
\begin{align}
	V (\omega \oplus 0) &= s_+ V, \label{V1} \\ 
	V (0 \oplus \omega) &= s_- V, \\ 
  v_x(s,\sigma):=V(w_x,0)(s,\sigma) &= v(s) e^{-is \sprod{\sigma}{x}},  \label{V2}
\end{align}
where $v(s) \coloneqq s w(s) \theta_+(s)$ and  $\theta_+(s) \coloneqq \rchi(s\ge 0)$ denotes the Heaviside function.
For the combined unitary mapping $\WW \coloneqq \Gamma(V)U^{*}: \FF(\hn) \otimes \FF(\hn) \to \FF(\h)$ it follows from \eqref{U1}-\eqref{V2} that 
\begin{align}
      \WW \Hex &= H\WW, \label{H1}\\
      \WW \big(\Hn \otimes 1\big) &= H_{+} \WW, \label{H2} \\ 
      \WW \big(1 \otimes H_\omega \big) &= H_{-} \WW, \label{H3} 
\end{align}
with operators in $\HH \coloneqq \HHel \otimes \FF(\h)$ defined by
\begin{align}
     H &\coloneqq \Hel \otimes 1 + 1 \otimes d\Gamma(s) + g\phi(v_x), \label{DEFINITION H}\\
     H_{+} &\coloneqq \Hel \otimes 1 + 1 \otimes d\Gamma(s_{+}) + g\phi(v_x), \label{DEFINITION H_+}\\ 
     H_- &\coloneqq 1 \otimes d\Gamma(s_-).  \label{DEFINITION H_-}
\end{align}
By an application of Nelson's commutator theorem, see \cite{JP1996}, the operator $H$ is essentially self-adjoint on any core of $\Hel \otimes 1 + 1 \otimes d\Gamma(|s|)$. By construction, the operators $H_+$ and $H_-$ commute and $H = H_{+}-H_{-}$. The operator $H_+$, being the unitary transform of $\Hn \otimes 1$, is bounded below with the same ground state energy $E$ as $\Hn$, and $e^{\eps|x|} f(H_+)$ is bounded for $f \in C_0^\infty(-\infty, \Sigma)$ and $\eps \ll 1$. 

The operator 
\begin{align}  \label{DEFINITION tH}
	\tH \coloneqq H\otimes 1 + 1 \otimes d\Gamma(s)
\end{align} 
in $\tHH \coloneqq \HH \otimes \FF(\h)$ is the unitary transform of $\Hext$, 
\begin{align}  \label{H4}
	(\WW \otimes \WW) \Hext = \tH (\WW \otimes \WW).
\end{align} 
In view of \eqref{H1}-\eqref{H4} and the identity $ \WW I_\mathrm{ex} = I \big(\WW\otimes \WW\big)$,
with $I$ the scattering identification on $\FF(\h) \otimes \FF(\h)$, we see that $\psi \otimes \Omega$ is a scattering state of $\Hex$, see \eqref{AC reformulated with Hex}, 
if and only if $\Psi \coloneqq \WW (\psi \otimes \Omega)$ is a scattering state of $H$. Explicitly this means that for every $\eps > 0$ there exist 
$\Phi \in \HH$, $\Lambda \in \FF_\mathrm{fin}(\h)$ and $T > 0$ such that for all $t > T$
\begin{align}  
\| e^{-i H t}\Psi - I e^{-i \tH t} \left(\rchi_{\{E\}}(H_+) \rchi_{\{ 0 \}}(H_-) \Phi \otimes \rchi_{\{ 0 \}}(H_-) \Lambda \right) \| < \eps. \label{AC reformulated with H}
\end{align} 
We used that the range of $\rchi_{\{ E \}}(\Hn)$ and $\rchi_{\{ 0 \}}(H_\omega)$ are spanned by $\psi_{gs}$ and $\Omega$, respectively. 
The following lemma expresses properties of $\psi$ in terms of $\Psi$.

\begin{lemma} \label{psi vs Psi}
Let $\psi \in \HHn$ and $\Psi = \WW (\psi \otimes \Omega)$. Let $N$ denote the number operator in both $\HHn$ and $\HH$. 
Then the following holds true:
\begin{enumerate}
\renewcommand{\labelenumi}{(\alph{enumi})}
\item The vector $\psi$ is a scattering state for $\Hn$ in the sense \eqref{acp} if and only if $\Psi$ is a scattering state for $H$ in the sense \eqref{AC reformulated with H}. 
\item For any Borel set $\Delta$, if $ \psi \in \ran \, \rchi_\Delta(\Hn)$ then $\Psi \in \ran \, \rchi_\Delta(H_+) \rchinull(H_-).$
\item For all $t$ and $m$ we have $\|\rchi(N \geq m) e^{-i \Hn t} \psi \| = \| \rchi(N \geq m) e^{-i H t}\Psi \|$.
\end{enumerate}
\end{lemma}

\begin{proof}
Statement (a) has been shown above. (b) If $\psi \in \ran \, \rchi_\Delta(\Hn)$ then
 $$\psi \otimes \Omega = \left(\rchi_\Delta(\Hn) \otimes \rchinull (H_\omega)\right) (\psi \otimes \Omega).$$ 
Upon applying $\WW$ on both sides and using \eqref{H2}, \eqref{H3}, we find
$\Psi = \rchi_\Delta(H_+) \rchinull(H_-) \Psi$. (c) From the trivial identity $(1 \otimes N) (e^{-i \Hn t}\psi \otimes  \Omega) = 0$, from
\begin{equation}\label{NW}
\WW \big(N \otimes 1 + 1 \otimes N \big) = N \WW
\end{equation}
and from \eqref{H2} it follows that 
\begin{align*}  
	\| \rchi(N \geq m) e^{-i \Hn t} \psi \| &= \| \rchi(N \otimes 1 +1\otimes N \geq m) \left( e^{-i \Hn t} \psi \otimes \Omega \right) \| \\ 
	 &= \| \rchi(N \geq m) e^{-i H_+ t}\Psi \| = \|\rchi(N \geq m) e^{-i H t}\Psi \|.
\end{align*} 
In the last equation, we used that $e^{-i H t} = e^{-i H_+ t}e^{i H_{-} t}$, and that $H_-$ commutes with $N$.
\end{proof}

To express Hypothesis (v) in terms of the expanded objects we need $A_+ \coloneqq \WW (A \otimes 1) \WW^{-1}$ with $A$ the second quantized dilation generator. We remark that 
$A_+ = 1 \otimes d\Gamma(a_+)$ with the  essentially self-adjoint operator $a_+$ given by 
\begin{align*}  
	a_+ &\coloneqq  \frac{1}{2}  \theta_+(s)\left(s r+ rs \right) \theta_+(s) \qquad (r = i \partial_s),\\ 
	D(a_+) &\coloneqq C_0^\infty(\R \backslash \{ 0 \} \times S^2) \subset \h.
\end{align*} 
This is a consequence of Stone's theorem, and the fact that $3$-dimensional dilations on the first summand in $\hn \oplus \hn$ are mapped by $V$ onto $1$-dimensional dilations in 
$\R_+$ on $\hh$. 

\begin{description}
\item[(V)] \label{V} For all $f\in C_0^{\infty}(E,\lambda)$, $s<1/2$, $g>0$ small enough, and $\Psi \in \HH$,
$$
      \|\rchi(N = 0) e^{-i H t} f(H_+) \Psi \| = O(t^{-s}) \| \expect{A_+} \Psi \| \qquad (t \to \infty).
$$
\end{description}

\begin{lemma} \label{vac-decay}
Hypothesis \textnormal{(v)} implies  \textnormal{(V)}.
\end{lemma}

\begin{proof}
Statement (v) clearly implies 
$$
 \| (\rchi(N = 0) e^{-i\Hn t} f(\Hn) \otimes 1) \psi \| = O(t^{-s}) \| (\expect{A} \otimes 1) \psi \|  \qquad (\psi \in \HHex).
$$
We estimate the expression on the left from below using $ \|\rchi(N \otimes 1 + 1 \otimes N = 0) \varphi \| \leq  \|(\rchi(N = 0) \otimes 1) \varphi \|$ for $\varphi \in \HHex$.
Next we transform the vectors in the norms by the unitary $\WW$. Using \eqref{NW} and the definition of $A_{+}$ we arrive at
\begin{align*}  
	\|\rchi(N = 0) e^{-i H_+ t} f(H_+) \Psi \| \leq O(t^{-s}) \| \expect{A_+} \Psi \|,
\end{align*} 
with $\Psi = \WW\psi$. Since $e^{-i H t}= e^{-i H_+ t}e^{i H_- t}$ and $H_-$ commutes with $N$ the assertion follows.
\end{proof}

In view of \Cref{psi vs Psi} and \Cref{vac-decay}, the sufficiency of condition \eqref{N-bound} in \Cref{N-implies-AC} 
will follow from 
\begin{theorem}\label{thm:AC}
Let $\mu > 1/2$, $\lambda \in (E, \Sigma)$ and $\Delta = [E, \lambda)$. Assume \textnormal{(V)} and that $g > 0$ is sufficiently small. If $\Psi \in  \ran \, \rchi_\Delta(H_+) \rchinull (H_-)$ 
with
\begin{equation}\label{Ex-N-bound}
     \rchi(N \geq m) e^{-i H t} \Psi \to 0 \quad (m, t \to \infty)
\end{equation}
then $\Psi$ is a scattering state of $H$ in the sense \eqref{AC reformulated with H}.
\end{theorem}

\noindent{\emph{Remark:} The set $\HH_N$ of vectors satisfying \eqref{Ex-N-bound} is a closed linear space, which is invariant under $e^{-i H t}$.
Moreover, since $H_{-}$ commutes with $H$ and $N$, $\HH_N$ is also invariant under the unitary groups generated by $H_{-}$ and by  $H_{+} = H+H_{-}$. This implies that $\HH_N$ is invariant under $g(H)$ and $g(H_{\pm})$ for arbitrary bounded Borel functions $g$.

The proof of \Cref{thm:AC} is given in \Cref{main thm proof}. It is based on the two main results from \Cref{sec:DS} and \Cref{sec:min-escape}. 
We conclude the present section with two auxiliary results. The first one, for suitable values of the parameters, expresses integrable decay of the particle-boson interaction.

\begin{lemma}\label{short-range}
Let $r$ denote the operator $i\partial_s$ in $\hh = L^2(\R \times S^2)$. 
Let $\varepsilon > 0$, $a > 0$ and $\alpha \in (0, 1]$. Then
$$
  \sup_{x \in \R^3} e^{-\varepsilon|x|} \|\rchi(r \geq a t^\alpha) v_x\| = O(t^{- \alpha(\mu + 3/2)}) \qquad (t \to \infty).
$$
\end{lemma}
\begin{proof} 
Let $\check{v}(r)$ and $\check{v}_x(r, \sigma)$ denote the inverse Fourier transform of $s \mapsto v(s)$ and $s \mapsto v_x(s, \sigma)$, respectively. Then $\check{v}_x(r, \sigma) = \check{v}(r - \sprod{\sigma}{x})$ and
\begin{align*}  
	\|\rchi(r \geq \varepsilon t^\alpha) v_x\|^2 &= \int_{|\sigma| = 1} dS(\sigma) \int_{r \geq a t^\alpha} | \check{v}_x(r, \sigma) |^2 dr \\
	&=  \int_{|\sigma| = 1} dS(\sigma) \int_{r \geq a t^\alpha - \sprod{\sigma}{x}} |\check{v}(r)|^2 dr \\ 
	&\leq 4 \pi \| \check{v} \|^2 \rchi(|x| > a t^\alpha/2) + 4 \pi \int_{r \geq a t^\alpha/2} |\check{v}(r)|^2 dr.
\end{align*} 
From \Cref{short-range lemma} it follows that the integral is $O(t^{-\alpha ( 2  \mu  + 3)})$. Hence
\begin{align*}  
	\sup_{x \in \R^3} e^{-\eps |x|} \|\rchi(r \geq \varepsilon t^\alpha) v_x\|  \leq O(e^{-\eps a t^\alpha/2}) + O(t^{- \alpha ( \mu   + 3/2)}).\qquad \qedhere
\end{align*} 
\end{proof}

\begin{proposition}[G\'erard's bound] \label{Gerard bound} 
Let $f \in C_0^\infty(\R)$.
\begin{enumerate}
\renewcommand{\labelenumi}{(\alph{enumi})}
\item The operator  $f(\Hn)$ leaves $D(N)$ invariant and for all $\psi \in f(\Hn) D(N)$ 
\begin{align}  
\sprod{e^{-i \Hn t}\psi}{N e^{-i \Hn t}\psi} &=O(t^{1/(2 + \mu)}) \,  \sprod{\psi}{(N + 1) \psi} \label{Gerard bound N for Hn} \qquad (t \to \infty).
\end{align} 
\item The operator  $f(H_+)$ leaves $D(N)$ invariant and for all $\Psi \in f(H_+) D(N)$ 
\begin{align}  
\sprod{e^{-i H t}\Psi}{N e^{-i H t}\Psi} &=O(t^{1/(2 + \mu)}) \,  \sprod{\Psi}{(N + 1) \Psi} \label{Gerard bound N} \qquad (t \to \infty).
\end{align} 
\end{enumerate}
\end{proposition}

\begin{proof}
(a) The bound \eqref{Gerard bound N for Hn} is established in the proof of Proposition 4.3 in \cite{Gerard2002}. See also Proposition A.1 in \cite{FauSig2014}. (b) In view of \eqref{NW}, and since $e^{-i \Hn t} \otimes 1$ commutes with $1 \otimes N$, statement (a) implies for all $\Psi \in f(H_+) D(N)$
\begin{align*}  
	\sprod{e^{-i H_+ t}\Psi}{N e^{-i H_+ t}\Psi} &=O(t^{1/(2 + \mu)}) \,  \sprod{\Psi}{(N + 1) \Psi}.
\end{align*} 
This proves the assertion since $e^{-i H t} = e^{i H_{-} t}e^{-i H_{+} t}$, where $e^{i H_- t}$ commutes with $N$. 
\end{proof}


\section{Deift-Simon wave operator} 
\label{sec:DS}

In this section, the Deift-Simon wave operator $W$ is constructed. To this end, it suffices that $\mu > -1/2$. We pick $\alpha$ such that $1/(\mu + 3/2) < \alpha \leq 1$ and keep it fixed throughout this section. 

Let $c, d$ be real numbers with $0 < c < d < 1$ and choose functions $j_0$ and $j_\infty$ in $C^\infty(\R; [0,1])$ such that
\begin{align}  
	j_0(r) &= 1\ \text{for}\ r \leq c, \quad j_\infty(r) = 1\ \text{for}\ r \geq d, \quad
	j_0^2 + j_\infty^2 = 1. \label{def-j}  
\end{align} 
We set $j_{0, t}(r) = j_0(r/t^\alpha)$ and $ j_{\infty, t}(r) = j_\infty(r/t^\alpha)$ with $r = i \frac{\partial}{\partial s}$ in $\hh = L^2(\R \times S^2)$. The operator 
$$j_t = j_{0,t} \oplus j_{\infty, t} :\hh \to \hh\oplus\hh$$ 
then satisfies  $j_t^{*}j_t = j_{0, t}^2 + j_{\infty, t}^2 = 1$ and hence $\uGamma(j_t):\HH\to\tHH$ has the property
$$
              \uGamma(j_t)^{*}  \uGamma(j_t) = 1.
$$
See \Cref{sec:Fock} for the definition of $\uGamma(j_t)$ and the necessary prerequisites on second quantization. 
The purpose of this section is to establish the following theorem.

\begin{theorem}\label{thm-W}
Let $\mu > -1/2$, $\lambda \in (E, \Sigma)$ and $\Delta = [E, \lambda)$. Let $\Psi \in  \ran \, \rchi_\Delta(H_{+})$  and assume that $\rchi(N \geq m) e^{-i H t}\Psi \to 0$ as $m, t \to \infty$. Then the limit
\begin{align}  
	W \Psi \coloneqq \lim_{t\to\infty} e^{i\tH t} \uGamma(j_t) e^{-iHt} \Psi
	\label{DEFINITION W}
\end{align} 
exists. Moreover, for every bounded Borel function $g:\R\to \C$
\begin{align}  \label{pull through}
                W g(H_{\pm}) \Psi = g(\tH_{\pm}) W\Psi,
\end{align} 
where $\tH_\pm = H_\pm \otimes 1 + 1 \otimes d\Gamma(s_\pm)$ in $\tHH$.
\end{theorem}
The assumption on the distribution of $N$ allows us, in \Cref{Bt lemma}, to introduce the resolvent $(N_t + \rho)^{-1}$ with $N_t$ counting the outwards moving bosons, see the figure below. This resolvent is essential for the proof of the propagation estimate \Cref{main propagation estimate}, as it makes the propagation observable uniformly bounded in time. To control its Heisenberg derivative we need propagation estimate \Cref{dn prop}, which we learned from G\'erard's paper \cite{Gerard2002}. \Cref{interaction integrable} and \Cref{ecc} are further preparations for the proof of the subsequent results.

The auxiliary observable $N_t$ is constructed as follows. Let $a, b$ be real numbers with $0 < a < b < c < d < 1$. Let $n \in C^\infty(\R; [0, 1])$ with $n(r) = 0 $ for $r \leq a$ and $n(r) = 1$ for $r \geq b$. We set $n_t(r) = n(r/t^\alpha)$ and $N_t = d\Gamma(n_t)$. The functions $j_0, j_\infty$ and $n$ are illustrated by the following figure. \\ 

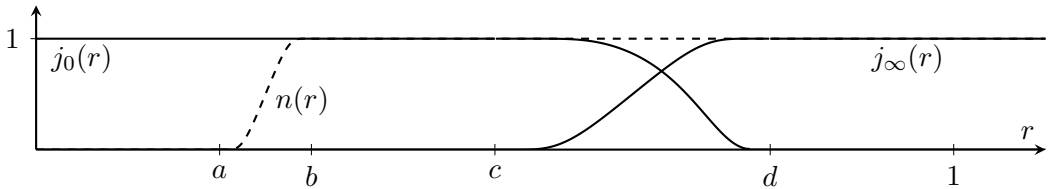
\begin{tikzpicture}
  \begin{axis}[
    axis lines=middle,
    xlabel={$r$},
    ylabel={},
    xmin=0, xmax=1.1,
    ymin=0, ymax=1.3,
    xtick={0.2,0.3,0.5,0.8,1},
    xticklabels={$a$, $b$, $c$, $d$, $1$},
    ytick={1},
    yticklabels={$1$},
    samples=400,
    domain=0:1.05,
    thick,
    smooth,
    no markers,
    width=15cm,
    height=3.5cm,
    grid=none,
    every axis plot/.append style={black},
    axis line style={black},
    tick style={black},
    ]
    
  \def\a{0.2}
  \def\b{0.3}
  \def\c{0.5}
  \def\d{0.8}
  \def\eps{0.001}

  \addplot [
    domain=0:\a,
    dashed,
  ] {0};

  \addplot [
    domain=\a + \eps:\b - \eps,
    dashed,
  ] 
  { 
    (exp(-1 / ((x - \a)/(\b - \a))) /
     (exp(-1 / ((x - \a)/(\b - \a))) + exp(-1 / (1 - (x - \a)/(\b - \a))))
    )
  };
  
  \addplot [
    domain=\b:1.1,
    dashed,
  ] {1};

  \addplot [
    domain=0:\c,
  ] {0};

  \addplot [
    domain=\c + \eps:\d - \eps,
  ] 
  { 
    (exp(-1 / ((x - \c)/(\d - \c))) /
     (exp(-1 / ((x - \c)/(\d - \c))) + exp(-1 / (1 - (x - \c)/(\d - \c))))
    )
  };

  \addplot [
    domain=\d:1.1,
  ] {1};
  
  \addplot [
    domain=0:\c,
  ] {1};  

  \addplot [
    domain=\c + \eps:\d - \eps,
  ] 
  { 
    sqrt(1 -
    (
      exp(-1 / ((x - \c)/(\d - \c))) /
      (exp(-1 / ((x - \c)/(\d - \c))) + exp(-1 / (1 - (x - \c)/(\d - \c))))
    )^2)
  };

  \addplot [
    domain=\d:1.1,
  ] {0};  

    \node[black] at (axis cs:0.05,0.82) {$j_0(r)$};
    \node[black] at (axis cs:0.95,0.82) {$j_\infty(r)$};
    \node[black] at (axis cs:0.29,0.44) {$n(r)$};

  \end{axis}
\end{tikzpicture}

  

\begin{lemma} \label{interaction integrable} 
Let $\eps > 0$. If $f_t:\R\to\C$ satisfies $0 \leq |f_t(r)| \leq n_t(r)$ then
\begin{align*}  
\| e^{-\eps |x|} \phi(f_t v_x) (N_t + 1)^{-1/2} \| = O(t^{- \alpha (\mu + 3/2)}) \qquad (t \to \infty).
\end{align*} 
\end{lemma}
\begin{proof}
By assumption, $f_t(r) \neq 0$ implies $n_t(r) \neq 0$. Let $g_t \coloneqq n_t^{-1/2} f_t$ in points where $f_t \neq 0$ and else $g_t = 0$. It follows that
\begin{align*}  
|f_t(r)| \leq n_t(r) \leq \rchi(r \geq a t^\alpha) \quad \text{and}\quad |g_t(r)| \leq n_t(r)^{1/2} \leq  \rchi(r \geq a t^\alpha).
\end{align*} 
Using $f_t = n_t^{1/2} g_t$ and $N_t = d\Gamma(n_t)$ it is straightforward to verify for $\eta \in \FF(\h)$, $x \in \R^3$,
\begin{align*}  
	\| a(f_t v_x) \eta \| \leq \| g_t v_x \| \sprod{\eta}{N_t \eta}^{1/2}.
\end{align*} 
Since $\| a^*(f_t v_x) \eta \| \leq \| a(f_t v_x) \eta \| + \| f_t v_x \| \| \eta \|$, it follows that
\begin{align*}  
	\| \phi(f_t v_x) \eta \| &\leq \big(2 \| g_t v_x \| + \| f_t v_x \| \big) \| (N_t + 1)^{1/2} \eta \| \\ 
	&\leq 3\| \rchi(r \geq a t^\alpha) v_x \| \| (N_t + 1)^{1/2} \eta \|. 
\end{align*} 
Hence, for $\Psi \in \HH$
\begin{align*}  
	\| e^{-\eps|x|} \phi(f_t v_x) \Psi \| \leq 3 \sup_{x \in \R^3} \left(e^{-\eps |x|} \| \rchi(r \geq a t^\alpha) v_x \|  \right) \| (N_t + 1)^{1/2} \Psi \|,
\end{align*} 
where the supremum is $O(t^{- \alpha(\mu + 3/2)})$ by \Cref{short-range}.
\end{proof}

\begin{lemma}\label{ecc} 
Let $g\in C_0^{\infty}(\R)$ and $f \in C_0^\infty(-\infty, \Sigma)$. Then for all $\Psi \in \HH$ 
\begin{align}  \label{W commute lemma}
     \Big(g(\tH_{\pm})\uGamma(j_t) - \uGamma(j_t) g(H_{\pm})\Big) f(H_+) \Psi_t \to 0 \qquad (t\to\infty), 
\end{align} 
where $\Psi_t = e^{-i H t} \Psi$ and $\tH_\pm = H_\pm \otimes 1 + 1 \otimes d\Gamma(s_\pm)$.
\end{lemma}

\begin{proof}
The following proof is inspired by Lemma 5.2 in \cite{FauSig2014}. Since 
$$B_t^\pm \coloneqq  \big(g(\tH_\pm)\uGamma(j_t) - \uGamma(j_t) g(H_\pm)\big)$$
 is bounded uniformly in $t$, it suffices to prove 
 \begin{align*}  
 	B_t^\pm f(H_+) \Psi_t \to 0 \qquad (t \to \infty), 
 \end{align*} 
 for $\Psi \in h(H_+) D(N)$ with $h \in C_0^\infty(-\infty, \Sigma)$ satisfying $f = f h$.
Below we will prove that 
\begin{align}  \label{commute op bound}
   \| B_t^\pm \, f(H_{+}) (N + 1)^{-1} \| = O(t^{-\alpha}).
\end{align} 
Now choose $\beta$ such that $1/(\mu + 2) < \beta < 1/(\mu + 3/2)$. Then, by \eqref{Gerard bound N}, 
\begin{align}  \label{Gerard Verteilung}
\| \rchi(N \geq t^\beta) \Psi_t \|^2 \leq t^{-\beta} \sprod{\Psi_t}{N \Psi_t} = O(t^{-\beta} t^{1/(\mu + 2)} ) \to 0.
\end{align} 
From \eqref{commute op bound} and \eqref{Gerard Verteilung} it follows that
\begin{align*}  
	\| B_t^\pm f(H_+) \Psi_t \| &\leq \| B_t^\pm \, f(H_+) \rchi(N \geq t^\beta) e^{-i H t} \Psi \| + \| B_t^\pm \, f(H_+) \rchi(N < t^\beta) e^{-i H t} \Psi \| \\ 
	&\leq o(1) + O(t^{- \alpha}) \| (N + 1) \rchi(N < t^\beta) \|  \\ 
	&= o(1) + O(t^{- \alpha}) O(t^{\beta}) \to 0 \qquad (t \to \infty), 
\end{align*} 
since $\beta < 1/(\mu + 3/2) < \alpha$.

It remains to prove \eqref{commute op bound}. We consider the ``$+$''-case only. The ``$-$''-case is similar and easier. 
The bound \eqref{commute op bound} will follow from  the HS-formula \eqref{DEFINITION HS FORMULA} after we have shown that 
\begin{equation}\label{RJ-commute}
     \| \Big(\tR_{+}(z)\uGamma(j_t) - \uGamma(j_t) R_{+}(z)\Big)  f(H_{+}) (N + 1)^{-1} \|  = O(t^{-\alpha}) \frac{1}{|\mathrm{Im} \, z|^3},
\end{equation}
where $\tR_{+}(z) = (z-\tH_{+})^{-1}$ and $R_{+}(z) = (z-H_{+})^{-1}$.  
We compute
\begin{align} \label{ecc aux 1}
     \lefteqn{\tR_{+}(z)\uGamma(j_t) - \uGamma(j_t) R_{+}(z)} \nonumber  \\
     &= \,  \tR_{+}(z)\big( \tH_{+} \uGamma(j_t) - \uGamma(j_t)H_{+}\big) R_{+}(z) \nonumber \\ 
     &= \,  \tR_{+}(z) \bigg( d\uGamma(j_t,[s_{+},j_t])  
      + g \big[ \phi((1-j_{0, t} )v_x)\otimes 1 - 1\otimes \phi(j_{\infty, t} v_x)\big]\uGamma(j_t) \bigg) R_{+}(z).
\end{align}
From $\| e^{\eps |x|} f(H_+)\|< \infty$ for $\eps \ll 1$ and \Cref{N f N inverse lemma} it follows that
\begin{align}  \label{ecc aux 2}
     \| (N+e^{\eps |x|} +1)R_{+}(z) f(H_{+}) (N + 1)^{-1}\| \le C \frac{1}{|\mathrm{Im} \, z|^2}.
\end{align} 
For the proof of \eqref{RJ-commute}, by \eqref{ecc aux 1} and \eqref{ecc aux 2}, it suffices to show that
\begin{align} \label{ecc aux 3}
	\|\big( d\uGamma(j_t,[s_{+},j_t])  + g \big[ \phi((1-j_{0, t} )v_x)\otimes1 - 1\otimes\phi(j_{\infty, t} v_x)\big]\uGamma(j_t) \big)& (N + e^{\eps |x|} + 1)^{-1}\| \nonumber \\ 
      &= O(t^{-\alpha}). 
\end{align}  
By \Cref{sqrt-Lap}, 
\begin{align}  \label{ecc aux 4}
 	\| d\uGamma(j_t,[s_{+},j_t]) (N+1)^{-1}\| \le \|[s_{+},j_t]\| =  O(t^{-\alpha}).
 \end{align} 
 Using $\uGamma(j_t) N = (N \otimes 1 + 1 \otimes N) \uGamma(j_t)$ we obtain
\begin{align} \label{ecc aux 5}
	\lefteqn{\| \big(\phi((1 - j_{0, t}) v_x) \otimes 1 \big) \uGamma(j_t) (N + e^{\eps |x|} + 1)^{-1} \|}  \nonumber \\ 
	&= \|\big(\phi((1 - j_{0, t}) v_x) \otimes 1 \big) (N \otimes 1 + 1 \otimes N + e^{\eps|x|} + 1)^{-1} \uGamma(j_t) \| \nonumber \\ 
	&\leq \|\phi((1 - j_{0, t}) v_x) (N + e^{\eps|x|} + 1)^{-1} \| \nonumber \\ 
	&\leq \|e^{-\eps|x|/2} \phi((1 - j_{0, t}) v_x) (N + 1)^{-1/2} \|= O(t^{-\alpha (\mu + 3/2)}), 
\end{align}
where in the last line we used  \Cref{interaction integrable}.
Similarly, the $\phi(j_{\infty, t} v_x)$-term in \eqref{ecc aux 3} is also $O(t^{-\alpha (\mu + 3/2)})$. 
Inequalities \eqref{ecc aux 4} and \eqref{ecc aux 5} imply \eqref{ecc aux 3} because $\mu + 3/2 > 1$.  
\end{proof}

The Heisenberg derivative $D A_t$  of operators $(A_t)_{t \in \R}$ in $\HH$ is defined by 
\begin{align*}  
	D A_t = [i H, A_t] + \partial_t A_t.
\end{align*} 
The corresponding free Heisenberg derivative is
\begin{align*}  
	D_0 A_t =  [i H_{g = 0}, A_t] + \partial_t A_t.
\end{align*} 
If  $A_t$ is an operator in $\tHH$, then $H$ is to be replaced with $\tH$. 
If $A_t$ is an operator from $\HH$ to $\tHH$, then $D A_t$ is defined by
\begin{align*}  
	D A_t = i (\tH A_t - A_t H) + \partial_t A_t.
\end{align*} 
Finally, for operators $(a_t)_{t \in \R}$ in $\hh$ we set $d a_t = [i s, a_t] + \partial_t a_t$.

\Cref{dn prop} and \Cref{main propagation estimate} below establish propagation estimates of the form
\begin{equation}\label{pe-strategy1}
   \int_1^{\infty} \big\|P(t)^{1/2}f(H_{+})\Psi_t\big\|^2\, dt \leq C\|\Psi\|^2,
\end{equation}
where $P(t)\geq 0$, $f \in C_0^\infty(-\infty, \Sigma)$, and $\Psi_t=e^{-iHt}\Psi$. The strategy of proof is to construct a suitable propagation observable 
$\phi(t)=\phi(t)^{*}$ that is bounded above uniformly in $t\geq 1$ and satisfies 
 \begin{equation}\label{pe-strategy2}
      D\phi(t) = P(t) + R(t).
 \end{equation}
 Here $R(t)$ is an \emph{integrable remainder} in the sense that
 \begin{equation}\label{pe-strategy3}
       \int_1^{\infty} \big|\sprod{f(H_{+})\Psi_t}{R(t)f(H_{+})\Psi_t}\big|\, dt \leq \const\|\Psi\|^2.
 \end{equation}
 Estimate \eqref{pe-strategy1} follows from \eqref{pe-strategy2} and \eqref{pe-strategy3} by integrating the expectation value of   \eqref{pe-strategy2} in the state
 $f(H_{+})\psi_t$.

The following proposition, with a different choice of $N_t$, agrees with Proposition 5.1(i)\footnote{The additional $1/t$ in  \cite{Gerard2002} is a typo.} in \cite{Gerard2002}. 
For completeness we give the short proof.

\begin{proposition} \label{dn prop}
Let $\rho >0$ and $f \in C_0^\infty(-\infty, \Sigma)$. Then $D_0 N_t = d\Gamma(d n_t) \geq 0$ and for all $\Psi \in \HH$
\begin{align}  
	\int_1^\infty \| (D_0 N_t)^{1/2} (N_t + \rho)^{-1} f(H_+) \Psi_t \|^2 dt \leq C \| \Psi \|^2.
	\label{dn bound}
\end{align} 
\end{proposition} 

\noindent \emph{Remark: } Let $\tilde{N_t} \coloneqq N_t \otimes 1 + 1 \otimes N_t$ in $\tHH$. Then for all $\Phi \in \tHH$
\begin{align*}  
	\int_1^\infty \|(D_0 \tilde{N_t})^{1/2}(\tilde{N_t} + \rho)^{-1}f(\tH_+)\Phi_t \|^2 dt \leq C \| \Phi \|^2,
\end{align*} 
where $\Phi_t = e^{-i \tH t}\Phi$. The proof is completely analogous to the proof of \eqref{dn bound}. 

\begin{proof}
For the proof of $d n_t \geq 0$ we compute
\begin{align*}  
	d n_t = [is, n_t] + \partial_t n_t = \frac{1}{t^\alpha} n'(r/t_\alpha) (1 - \alpha r/t).
\end{align*} 
We have $n' \geq 0$ and $(1 - \alpha r/t) \geq (1 - \alpha r/t^\alpha) \geq (1 - \alpha b) > 0$ for $t \geq 1$ and $r/t^\alpha$ in the support of $n'$. 

For the proof of \eqref{dn bound} we define $\phi(t) \coloneqq -(N_t + \rho)^{-1}$. Then $-1/\rho \leq \phi(t) \leq 0$ and
\begin{align*}  
	D \phi(t) &= (N_t + \rho)^{-1} (D N_t) (N_t + \rho)^{-1} \\ 
	&= (N_t + \rho)^{-1} d\Gamma(d n_t) (N_t + \rho)^{-1} - g (N_t + \rho)^{-1} \phi(i n_t v_x) (N_t + \rho)^{-1}.
\end{align*} 
Since $e^{\eps |x|} f(H_+)$ is bounded for small $\eps $, we see, by \Cref{interaction integrable} and the assumption $\alpha(\mu+3/2)>1$, that the $\phi(i n_t v_x)$-term is an integrable remainder in the sense of \eqref{pe-strategy3}. Since $\phi(t)$ is bounded uniformly in $t\geq 1$, the bound \eqref{dn bound} follows. 
\end{proof}


The following propagation estimate is our main technical innovation.

\begin{proposition}\label{main propagation estimate}
Let $\rho >0$ and $f \in C_0^\infty(-\infty, \Sigma)$. Then for all $\Psi \in \HH$
\begin{align}  \label{X bound}
	\int_1^\infty \| d\Gamma(\rchi_{[c, d]}(r/t^\alpha))^{1/2} (N_t + \rho)^{-1} f(H_+) \Psi_t \|^2 \frac{dt}{t^\alpha} \leq C \| \Psi \|^2.
\end{align} 
\end{proposition}
\noindent \emph{Remark: } Let $X_t \coloneqq d\Gamma(\rchi_{[c, d]}(r/t^\alpha))$ and $\tilde{X_t} \coloneqq X_t \otimes 1 + 1 \otimes X_t$ in $\tHH$. Then for all $\Phi \in \tHH$
\begin{align*}  
\int_1^\infty \| \tilde{X_t}^{1/2}(\tilde{N_t} + \rho)^{-1}f(\tH_+) \Phi_t \|^2 \frac{dt}{t^\alpha} \leq C \| \Phi \|^2,
\end{align*} 
where $\Phi_t = e^{-i \tH t} \Phi$. The proof is completely analogous to the proof of \eqref{X bound}. 

\begin{proof} 
Choose numbers $c', d'$ such that $b < c' < c < d < d' < 1$ and pick $h\in C_0^{\infty}(\R)$ with $\rchi_{[c,d]}\le h\le \rchi_{[c',d']}$. Let 
$$
      \tilde{h}(r) := \int_0^r h(u)\, du.
$$
Then  $0 \leq \tilde{h} \le (d'-c')n$. Let  $\tilde{h}_t(r) \coloneqq \tilde{h}(r/t^\alpha)$ and recall that $n_t(r)=n(r/t^\alpha)$.
With the short-hand $R_t \coloneqq (N_t + \rho)^{-1}$ we define the propagation observable 
\begin{align*}  
	\phi(t) \coloneqq R_td\Gamma(\tilde{h}_t)  R_t.
\end{align*} 
From $0 \leq \tilde{h}_t \leq (d' - c') n_t$ and $R_t \, N_t \, R_t \le 1/\rho$ it follows that 
\begin{align}  \label{uniform bound on phi}
	0 \leq \phi(t) \le (d' - c')/\rho.
\end{align} 
We have
\begin{align*}  
	D \phi(t) = (D R_t) d\Gamma(\tilde{h}_t) R_t + R_t \big(D d\Gamma(\tilde{h}_t)\big)R_t + R_t d\Gamma(\tilde{h}_t)(D R_t).
\end{align*} 
We claim that the first and the third terms, both containing $D R_t$, are integrable remainders in the sense of \eqref{pe-strategy3}. 
It suffices to prove this for the first one. Using $DR_t=-R_t (DN_t) R_t$ we get
\begin{align*}  
	(D R_t) d\Gamma(\tilde{h}_t) R_t  =& - R_t d\Gamma(d n_t) R_t d\Gamma(\tilde{h}_t) R_t \\
	&+  g R_t \phi(i n_t v_x) R_t d\Gamma(\tilde{h}_t) R_t. 
\end{align*} 
The $\phi(i n_t v_x)$-term is an integrable remainder thanks to the exponential decay on $\ran f(H_+)$ in combination with \Cref{interaction integrable} and $\alpha(\mu+3/2)>1$. For the term involving $R_t d\Gamma(d n_t) R_t$ we notice that $R_t, d\Gamma(\tilde{h}_t)$ and $d\Gamma(d n_t)$ commute. So $\tilde{h}_t \leq (d' - c') n_t$ implies $d\Gamma(\tilde{h}_t) R_t \leq (d' - c')$ and hence 
\begin{align*}  
	0 \leq R_t d\Gamma(d n_t) R_t d\Gamma(\tilde{h}_t) R_t &= R_t d\Gamma(d n_t)^{1/2} d\Gamma(\tilde{h}_t) R_t d\Gamma(d n_t)^{1/2} R_t \\ 
	&\leq (d' - c') R_t d\Gamma(d n_t) R_t.
\end{align*} 
This is an integrable remainder thanks to \Cref{dn prop}. We conclude that 
\begin{align*}  
	D \phi(t) = R_t (D d\Gamma(\tilde{h}_t))R_t + (\text{integrable}).
\end{align*} 
Next, we compute
\begin{align*}  
	D d\Gamma(\tilde{h}_t) &=  d\Gamma(d \tilde{h}_t ) -  g \phi(i \tilde{h}_t v_x),
\end{align*}  
where
$$
   d\tilde{h}_t = [is,\tilde{h}_t] + \partial_t \tilde{h}_t = \frac{1}{t^\alpha} h(r/t^{\alpha})\Big(1 - \frac{\alpha r}{t}\Big).
$$
Since $r/t^{\alpha}\le d'$ on the support of $h(r/t^{\alpha})$, $t\ge t^{\alpha}$ for $t\ge 1$, and $h \ge \rchi_{[c, d]}$ it follows that
\begin{align*}  
      d\tilde{h}_t \ge  (1-\alpha d')  \rchi_{[c, d]}(r/t^\alpha) \frac{1}{t^\alpha},
\end{align*}
where $(1 - \alpha d') > 0$.  We conclude that
\begin{align*}  
	D \phi(t) \geq (1 - \alpha d') R_t   \rchi_{[c, d]}(r/t^\alpha) \frac{1}{t^\alpha}  R_t + (\text{integrable}),
\end{align*} 
where we applied \Cref{interaction integrable} to the $\phi(i \tilde{h}_t v_x)$-term. By the remarks preceding \Cref{dn prop}, this proves the theorem. 
\end{proof}

\begin{lemma} \label{Bt lemma}
Let $(B_t)_{t \in \R}$ be a family of uniformly bounded operators. Let $\Psi \in \HH$ and $\Psi_t = e^{-i H t} \Psi$. 
Suppose $\rchi(N \geq m) \Psi_t \to 0$ as $m, t \to \infty$ and that for each $\rho>0$ the limit
\begin{equation*}
     \lim_{t \to \infty} B_t (N_t + \rho)^{-2} \Psi_t
\end{equation*}
exists. Then $\displaystyle \lim_{t \to \infty} B_t \Psi_t$ exists.
\end{lemma}

\begin{proof}
Let $C_{t, \rho} \coloneqq 1 - \rho^2 (N_t + \rho)^{-2}$. Then for any $m\in \N$,
\begin{equation}\label{Bt-L1}
	B_t \Psi_t - B_t \rho^2 (N_t + \rho)^{-2} \Psi_t =  B_t C_{t, \rho} \rchi(N \geq m) \psi_t + B_t C_{t, \rho} \rchi(N < m) \Psi_t.
\end{equation}
By hypothesis and since $\| C_{t, \rho} \| \leq 2$, the first term can be made arbitrarily small by choosing $m$ and $t$ large.  
Concerning the second term of \eqref{Bt-L1} we note that, for fixed $m\in \N$, since $0\leq N_t\leq N$,
\begin{align*}  
0 &\leq C_{t, \rho} \rchi(N < m) = (N_t^2 + 2 \rho N_t) (N_t + \rho)^{-2} \rchi(N < m) \\
&\leq (m^2 + 2 \rho m) / \rho^2 \to 0 \qquad (\rho \to \infty).
\end{align*} 
This shows that the norm of \eqref{Bt-L1} can be made smaller than any $\eps>0$ by choosing first $m,t$ and then $\rho$ sufficiently large. This 
is sufficient to check the Cauchy condition for $t\mapsto B_t \Psi_t$ given the existence of $\lim_{t\to\infty} B_t \rho^2 (N_t + \rho)^{-2} \Psi_t$.
\end{proof}

\begin{proof}[Proof of \Cref{thm-W}] The following proof is inspired by \cite{FauSig2014, Gerard2002}.
Since $\Psi \in \ran \,  \rchi_\Delta(H_+)$, we may pick  $f \in C_0^\infty(-\infty, \Sigma)$, real-valued, such that $\Psi = f(H_+) \Psi$. For existence of $W\Psi = Wf(H_+)\Psi$ it suffices, by \Cref{ecc},
to prove the existence of 
\begin{align*}  
	\lim_{t \to \infty} e^{i \tH t} f(\tH_+) \uGamma(j_t) f(H_+) e^{-i H t} \Psi.
\end{align*} 
In view of \Cref{Bt lemma} and the assumption $\rchi(N \geq m) e^{-i H t} \psi \to 0$ as $m, t \to \infty$, the above limit exists provided  
\begin{align}  \label{N regularized W}
	\lim_{t \to \infty} e^{i \tH t} f(\tH_+) \uGamma(j_t) R_t^2 f(H_+) e^{-i H t} \Psi, 
\end{align} 
exists, where $R_t \coloneqq (N_t + \rho)^{-1}$ and $\rho > 0$. We now prove existence of \eqref{N regularized W} using \Cref{abstract cauchy} in combination with the propagations estimates \Cref{dn prop} and \Cref{main propagation estimate}.

With $\tilde{R_t} \coloneqq (\tilde{N_t} + \rho)^{-1}$, $\tilde{N}_t \coloneqq N_t \otimes 1 + 1 \otimes N_t$ we have $\uGamma(j_t) R_t= \tilde{R_t} \uGamma(j_t)$. In the weak sense,
\begin{align}  
	\frac{d}{dt}  &e^{i \tH t} f(\tH_+) \tilde{R_t} \uGamma(j_t) R_t f(H_+) e^{-i H t} \nonumber \\  
	&= e^{i \tH t} f(\tH_+) \left[ (D \tilde{R_t}) \uGamma(j_t) R_t + \tilde{R_t} (D \uGamma(j_t)) R_t + \tilde{R_t} \uGamma(j_t) (D R_t)  \right] f(H_+) e^{-i H t}, \label{eqW1}
\end{align} 
where
\begin{align}  
	D \tR_t &= - \tR_t (D \tilde{N}_t )\tR_t \nonumber \\ 
	&= - \tR_t \left(- g \phi(i n_t v_x) \otimes 1 +  D_0 \tilde{N}_t  \right) \tR_t, \label{eqW2}\\
\text{and}\qquad D R_t &= - R_t (D N_t )R_t \nonumber \\ 
	&= - R_t \left(- g \phi(i n_t v_x)  +  D_0 N_t \right) R_t, \label{eqW3}
\end{align} 
with the free Heisenberg derivatives 
\begin{align*}  
D_0 N_t &= d\Gamma(d n_t), \\ 
D_0 \tilde{N}_ t &= d\Gamma(d n_t) \otimes 1 + 1 \otimes d\Gamma(d n_t).
\end{align*} 
Moreover, 
\begin{align}  \label{eqW4}
	D \uGamma(j_t) = d\uGamma(j_t , dj_t) + i g \big[\phi((1-j_{0, t})v_x)\otimes 1 - 1\otimes \phi(j_{\infty, t} v_x)\big] \uGamma(j_t).
\end{align} 
Since $e^{\eps|x|} f(H_+)$ is bounded for small $\eps$, \Cref{interaction integrable} implies
that all interaction terms in \eqref{eqW2}-\eqref{eqW4} give integrable contributions to \eqref{eqW1} in the sense of \Cref{abstract cauchy}.
Since $d n_t \geq 0$ commutes with $j_t$, we have $\uGamma(j_t) (D_0 N_t) = (D_0 \tilde{N_t}) \uGamma(j_t)$ and hence
\begin{align*}  
	\uGamma(j_t) (D_0 N_t)^{1/2} = (D_0 \tilde{N_t})^{1/2} \uGamma(j_t).
\end{align*} 
We conclude that
\begin{align}  
\frac{d}{dt}  &e^{i \tH t} f(\tH_+) \tilde{R_t} \uGamma(j_t) R_t f(H_+) e^{-i H t}  \nonumber \\ &= e^{i \tH t} f(\tH_+) \bigg[ - 2 \tR_t ( D_0 \tilde{N}_t)^{1/2} \uGamma(j_t) R_t (D_0 N_t)^{1/2} R_t  
+ \tR_t d\Gamma(j_t, d j_t) R_t \bigg] f(H_+) e^{-i H t} \nonumber \\ &\hspace{110mm}+ (\text{integrable}). \label{eqW5}
\end{align} 
To check the conditions of \Cref{abstract cauchy} we apply \eqref{eqW5} to a vector $\Psi \in \HH$, take the inner product with $\Phi \in \tHH$, and then estimate term by term. For the first term on the right-hand side we have the bound
\begin{align*}  
	2 \, \| \uGamma(j_t) R_t \| \,  \| (D_0 \tilde{N}_t)^{1/2} \tR_t f(\tH_+) \Phi_t \| \,  \| (D_0 N_t)^{1/2} R_t f(H_+) \Psi_t \|.
\end{align*} 
By \Cref{dn prop} this bound satisfies the integrability conditions of \Cref{abstract cauchy}. For the inner product of the second term of \eqref{eqW5} we obtain, using \Cref{lm:udGamma},
\begin{align*}  
	\lefteqn{|\sprod{\Phi_t}{f(\tH_+) \tR_t d\uGamma(j_t, d j_t) R_t f(H_+) \Psi_t} |} \\ 
	&\leq  \| (d\Gamma(|dj_{0,t}|)^{1/2} \otimes 1) \tR_t f(\tH_+) \Phi_t \| \, \| d\Gamma(|dj_{0,t}|)^{1/2} R_t f(H_+) \Psi_t \|  \\ 
	&+ \| (1 \otimes d\Gamma(|dj_{\infty,t}|)^{1/2}) \tR_t f(\tH_+) \Phi_t \| \, \| d\Gamma(|dj_{\infty,t}|)^{1/2} R_t f(H_+) \Psi_t \|, 
\end{align*} 
where $|dj_{0,t}|$ and $|dj_{\infty,t}|$ are given by
\begin{align*}
     |dj_{0,t}(r)| &= \frac{1}{t^\alpha}|j_0'(r/t^\alpha)| (1-\frac{r\alpha}{t}), \\
     |dj_{\infty,t}(r)| &= \frac{1}{t^\alpha}|j_\infty'(r/t^\alpha)| (1-\frac{r\alpha}{t}).
\end{align*}
Since $r/t^\alpha \in [c,d]$ for $r/t^\alpha$ in the support of $j_0'$ or $j_\infty'$, we see that both operators are bounded above by a multiple of $t^{-\alpha}\rchi_{[c,d]}(r/t^\alpha)$. 
By \Cref{main propagation estimate} this bound also satisfies the integrability conditions of \Cref{abstract cauchy}. This concludes the proof of existence of \eqref{N regularized W} and hence of 
\eqref{DEFINITION W}.

It remains to prove \eqref{pull through} for all $g\in \mathcal{B}(\R)$, the set of bounded Borel functions. Existence of $Wg(H_{+})\Psi$ follows from the fact that $g(H_{+})\Psi$, by the remark following \Cref{thm:AC}, shares the relevant properties of $\Psi$. Let $\mathcal{E}\subset\mathcal{B}(\R)$ denote the subset for which \eqref{pull through} is true. Then $C_0^\infty(\R) \subset \mathcal{E}$, by  \Cref{ecc}, and $\mathcal{E}$ is closed under pointwise limits of uniformly bounded functions. Indeed, if $g_n\in \mathcal{E}$, $\sup_{n \in \N, \, x \in \R} |g_n(x)| < \infty$ and $g_n(x)\to g(x)$, then 
$$
     W g(H_\pm) \Psi = \lim_{n \to \infty} W g_n(H_\pm) \Psi = \lim_{n \to \infty} g_n(\tH_\pm) W \Psi = g(\tH_\pm) W \Psi.
$$
It follows that $\mathcal{E} = \mathcal{B}(\R)$.
\end{proof}


\section{Minimal escape property} 
\label{sec:min-escape}

This section is devoted to the minimal escape property, \Cref{min-escape}. Our proofs are inspired by the proofs of analogous results from \cite{FauSig2014, HSS1999}. 

\begin{theorem}\label{min-escape}
Let $\mu > 1/2$ and $\lambda \in (E, \Sigma)$. Assume that \textnormal{(V)} holds and $g \ll 1$. 
Let $\alpha \in (0,  \frac{1 + \mu}{2 + \mu})$. Then for all $f \in C_0^\infty(E, \lambda)$
$$
	\Gamma(\rchi(r \leq t^\alpha)) e^{-i H t} f(H_+) \xrightarrow{\hspace{1.5mm} s \hspace{1.5mm}} 0 \qquad (t \to \infty).
$$
\end{theorem}
\Cref{min-escape} will easily follow from \Cref{min-esc}, below. The proof is given at the end of the section. 

Let $\theta_{-}$ denote the characteristic function of $(-\infty,0]$ and let $\rchi\in C^\infty(\R,[0,1])$ be a smooth version of $\theta_{-}$ with $\supp(\rchi)\subset (-\infty,0]$ and $\rchi=1$ on $(-\infty,-\eps]$ with $\eps>0$ chosen later. Let $\rchi' = -\xi^2$ with $\xi \in C_0^{\infty}(\R)$.
\begin{lemma}\label{min-esc}
Let $\rchi$ be as described above. Let $c \in (0, 1)$ and assume the hypotheses of \Cref{min-escape}. Then there exists a dense subspace $D \subset \HH$ such that for all  $\Psi \in  f(H_{+}) D$ with $f \in C_0^\infty(E, \lambda)$ 
\begin{align}  \label{min-esc assertion}
    \sprod{\Psi_t}{\rchi\left(B/t-c \right)\Psi_t} = O(t^{\nu - 1}) \qquad (t\to\infty),
\end{align}
where $\Psi_t = e^{-i H t} \Psi$, $B = d\Gamma(r)$ and $\nu = 1/(2+\mu)$.
\end{lemma}

\begin{proof}
The following proof is inspired by \cite{FauSig2014} and \cite{HSS1999}. 

Let $\hh_0 \coloneqq C_0^\infty(\R \backslash \{ 0 \} \times S^2)$. Then $\hh_0$ is dense in  $\hh$ and  $D \coloneqq \HHel \otimes \FF_\mathrm{fin}(\hh_0)$ is dense in $\HH$. The subspace $D$ is contained in $D(H_-), D(A_+)$, $D(B)$ and $D(N)$.  Let $\Psi = f(H_{+}) \Phi$ with $\Phi \in D$. Then $\Psi \in D(H)$ and, by \Cref{invariance of domains}, $\Psi \in D(B) \cap D(N)$.

Let $B_T \coloneqq (B - ct)/T$ and $\phi_T(t) \coloneqq \rchi(B_T)$. Our strategy is to first estimate $\sprod{\Psi_t}{\phi_T(t) \Psi_t}$ for fixed $T$ and then choose $T = t$ to obtain the bound \eqref{min-esc assertion}. To that end, we define a residual operator $R$ by the "chain rule" equation 
\begin{equation} \label{min vel commutator}
 [iH,\rchi(B_T)] = - \xi(B_T) [iH,B_T] \xi(B_T) + R,
 \end{equation}
where $[i H, B_T] = \frac{1}{T} (N - g \phi(i rv_x))$. Then, using the abbreviation $\xi = \xi(B_T)$, we compute the Heisenberg derivative 
\begin{align} \label{min-esc aux 1}
     D\phi_T &= [iH,\phi_T] +\partial_t\phi_T \nonumber \\
                  &= - \xi [iH,B_T] \xi + R +  \frac{c}{T}\xi^2 \nonumber \\
                  &= - \frac{1}{T} \xi\Big(N-g\phi(ir v_x) -c\Big)\xi +R.
\end{align}
We have $N-g\phi(ir v_x) = (1-g)N + g(N-\phi(ir v_x) )$, where $N \geq 1 - \rchi(N = 0)$ and $N-\phi(ir v_x) \ge -\|rv_x\|^2 \ge - C \expect{x}^2$. Hence
\begin{align*}  
	N-g\phi(ir v_x)  \geq 1 - \rchi(N = 0) + O(g) - C g \expect{x}^2. 
\end{align*} 
Since $\Psi_t = \rchi(H_+ \leq \lambda) \Psi_t$, since $\expect{x} \rchi(H_+ \leq \lambda)$ is bounded, and since $\expect{x}$ commutes with $\xi$, it follows that the expectation w.r.t $\Psi_t$ satisfies, for $g$ small enough depending on $c$,
\begin{align} \label{min-esc aux 2}
	  \xi\Big(N-g\phi(ir v_x) -c\Big)\xi \geq - \xi \rchi(N = 0) \xi + \xi (1 - O(g) - c) \xi  \geq - \| \xi \|^2 \rchi(N = 0).
\end{align}
From \eqref{min-esc aux 1} and \eqref{min-esc aux 2} it follows that 
\begin{align} \label{min-esc aux 3}
     \sprod{\Psi_t}{D\phi_T(t) \Psi_t}  \leq \frac{\|\xi\|^2}{T} \, \sprod{\Psi_t}{\rchi(N = 0)\Psi_t} + \sprod{\Psi_t}{R \Psi_t}.
\end{align}
Let $s  \coloneqq (1 - \nu)/2 < 1/2$. From Hypothesis (V) it follows that
\begin{align} \label{min-esc aux 4}
      \sprod{\Psi_t}{\rchi(N = 0) \Psi_t} =  O(t^{-2s}) \| \expect{A_{+}} \Phi \|^2 = O(t^{\nu - 1}).
\end{align}
Below we will prove that $R$, defined by \eqref{min vel commutator}, satisfies the bound 
 \begin{equation} \label{min vel error}
  \sprod{\Psi_t}{R\Psi_t} \le \frac{\text{const.}}{T^2}\sprod{\Psi_t}{(N + 1) \Psi_t}.
 \end{equation}
Combined with  G\'erard's bound $\sprod{\Psi_t}{(N + 1)\Psi_t} = O(t^\nu)$, see  \eqref{Gerard bound N}, it follows that 
\begin{align} \label{min-esc aux 5}
     \sprod{\Psi_t}{R\Psi_t} = O(t^\nu)/T^2.
\end{align} 
Integrating the upper bounds \eqref{min-esc aux 3}, \eqref{min-esc aux 4} and \eqref{min-esc aux 5} we find
\begin{align} \label{min-esc final 1}
    \sprod{\Psi_t}{\phi_T(t)\Psi_t} \le \sprod{\Psi}{\phi_T(0)\Psi} + O(t^{\nu}/T) + O(t^{1+\nu}/T^2).
\end{align}
By construction of $\rchi$, $\rchi(x)\leq \text{const.}|x|$. So 
\begin{align} \label{min-esc final 2}
     \sprod{\Psi}{\phi_T(0)\Psi} = \sprod{\Psi}{\rchi(B/T)\Psi} \leq \text{const.} \sprod{\Psi}{|B|\Psi}/T. 
\end{align}
The bound \eqref{min-esc assertion} follows from \eqref{min-esc final 1} and \eqref{min-esc final 2} with the choice $T = t$. 

It remains to verify the bound \eqref{min vel error} for $R$. By a kind of chain rule (see below), by the identity $\rchi' = - \xi^2$ and by the IMS formula, 
 \begin{align}  
 	[i H, \rchi(B_T)] &= \frac{1}{2} \rchi' [i H, B_T] + \frac{1}{2} [i H, B_T] \rchi' + R_1 \label{eqme1} \\ 
				&= - \xi [i H, B_T] \xi + R_2 + R_1, \nonumber 
 \end{align} 
 where 
 \begin{align*}  
 	R_2 = - \frac{1}{2} [[i H, B_T], \xi], \xi] =  \frac{1}{2 T} [[\phi(i r v_x), \xi], \xi].
 \end{align*} 
 So $R = R_1 + R_2$, where $R_1$ is defined by \eqref{eqme1} and estimated below. 
 
 First, we estimate $R_2$. To this end, note that $v(s) = s^{\mu + 1} \theta_+(s) \zeta(s)$ has weak derivatives $v', v''$ in $L^2(\R)$ because $\mu > 1/2$ by assumption. We conclude that 
\begin{align}  
	\sup_{x \in \R^3} \expect{x}^{-2} \| r^2 v_x \| &< \infty. \label{sup r2}
\end{align} 
From the Helffer-Sj\"ostrand formula for $\xi(B_T)$, see \eqref{DEFINITION HS FORMULA}, it follows that
\begin{align*}  
[\phi(i r v_x), \xi] = \frac{-i}{T} \int  (z - B_T)^{-1} \phi(r^2 v_x) (z - B_T)^{-1} d \tilde{\xi}(z),
\end{align*} 
where the extension $\tilde{\xi}$ of $\xi$ is chosen such that $|\partial_{\overline{z}} \tilde{\xi}(z)|/|\mathrm{Im} z|^2 $ is integrable. It follows that 
\begin{align*}  
|\sprod{\Psi_t}{R_2 \Psi_t}| &\leq \frac{1}{T} | \sprod{\xi \Psi_t}{[\phi(i r v_x), \xi] \Psi_t} | \\ 
&\leq \frac{\text{const.}}{T^2} \int \frac{1}{|\mathrm{Im} z|^2} \, \|\Psi_t\| \, \| (N + 1)^{1/2} \Psi_t\| \, |d\tilde{\xi}(z)| \\ 
&\leq \frac{\text{const.}}{T^2}\sprod{\Psi_t}{(N + 1) \Psi_t},
\end{align*} 
where we used that $\expect{x}^2 \rchi(H_+ \leq \lambda)$ is bounded and that \eqref{sup r2} implies 
\begin{align} \label{min-esc aux 6}
\| \expect{x}^{-2}\phi(r^2 v_x) (N + 1)^{-1/2} \| < \infty.
\end{align} 

To estimate $R_1$ in \eqref{eqme1} one is tempted to use the Helffer-Sj\"ostrand formula for $\rchi(B_T)$, but this is not directly possible because $\rchi$ is not compactly supported. We therefore make an approximation argument with the help of a compactly supported cutoff function: Let $\eta \in C_0^\infty(\R)$ with $\eta= 1$ in a neighborhood of $0$. We set $\eta_\varepsilon(x) \coloneqq \eta(\varepsilon x)$ and $\rchi_\varepsilon \coloneqq \eta_\varepsilon \rchi$. We are going to prove that
\begin{align}  \label{first step eps}
[i H, \rchi_\varepsilon(B_T)] = \frac{1}{2} \rchi_\varepsilon'  \, [i H, B_T] + \frac{1}{2}  [i H, B_T] \rchi_\varepsilon'  + R_{1, \varepsilon}
\end{align} 
with an operator $R_{1, \varepsilon}$ satisfying \eqref{min vel error} \emph{uniformly} in $\varepsilon$. Since $\rchi_\varepsilon \to \rchi$ and $(\rchi_\varepsilon)' = \rchi' \eta_\varepsilon + O(\varepsilon) \to \rchi'$ strongly, as $\varepsilon \to 0$, this will conclude the proof.  By \Cref{analytic extension} there exists an almost analytic extension $\tilde{\rchi}_\varepsilon$ of $\rchi_\varepsilon$ such that uniformly in $\varepsilon$
\begin{align}  
	\supp \tilde{\rchi}_\eps &\subset \{z \in \C \, | \, | y | \leq 2 \expect{x} \}, \label{uniform extension bound 1} \\ 
        | \partial_{\bar{z}}\tilde{\rchi}_\varepsilon(z) | &\leq  \text{const.} \,  \expect{x}^{-1 - 3} | y|^3, \quad z = x + i y. \label{uniform extension bound 2}
\end{align} 
From the HS-formula \eqref{DEFINITION HS FORMULA} for $\rchi_\eps(B_T)$ it follows that
\begin{align*}  
[i H, \rchi_\varepsilon(B_T)] &= \int (z - B_T)^{-1} [i H, B_T] (z - B_T)^{-1} d\tilde{\rchi_\varepsilon}(z).
\end{align*} 
By commuting $(z - B_T)^{-1}$ once to the left and once to the right of $[iH, B_T]$ we obtain 
\begin{align*}
	[i H, \rchi_\varepsilon(B_T)] = \frac{1}{2} \rchi_\varepsilon'  \, [i H, B_T] + \frac{1}{2}  [i H, B_T] \rchi_\varepsilon'  + R_{1, \varepsilon}
\end{align*}
with 
\begin{align*}  
R_{1, \varepsilon} = \frac{1}{2} \int (z - B_T)^{-2} [[i H, B_T], B_T] (z - B_T)^{-1} d\tilde{\rchi_\varepsilon}(z) + \text{h.c.}
\end{align*} 
From $[[i H, B_T], B_T] = \frac{1}{T^2} i \phi(r^2 v_x)$, $\| \expect{x}^2 \rchi(H_+ \leq \lambda)\|< \infty$, \eqref{min-esc aux 6}, \eqref{uniform extension bound 1} and \eqref{uniform extension bound 2} we find
\begin{align*}  
| \sprod{\Psi_t}{R_{1, \varepsilon} \Psi_t} | &\leq \frac{\text{const.}}{T^2}  \int_{-\infty}^\infty \hspace{-3mm} dx \int_{-2 \expect{x}}^{2 \expect{x}} \hspace{-3mm} dy \,  \expect{ x }^{-1 - 3} |y|^3 \frac{1}{|y|^3} \| \Psi_t \| \cdot \| (N + 1)^{1/2} \Psi_t \| \\ 
&\leq \frac{\text{const.}}{T^2} \, \sprod{\Psi_t}{(N + 1) \Psi_t}. \qedhere
\end{align*} 
\end{proof}

\begin{proof}[Proof of \Cref{min-escape}]
Pick $\alpha' \in  (0,  \frac{1 + \mu}{2 + \mu})$ with $\alpha' > \alpha$ and $c' \in (0, 1)$. Since $\rchi(r \leq t^\alpha) \leq \rchi(r \leq c' t^{\alpha'})$ for large $t$, it suffices to show that 
$$
	\Gamma(\rchi(r \leq c' t^{\alpha'})) e^{-i H t} f(H_+) \xrightarrow{\hspace{1.5mm} s \hspace{1.5mm}} 0.
$$
Let $\theta_- = \rchi_{(-\infty, 0]}$. For the above statement it suffices to prove that 
\begin{align*}  
	\|\Gamma(\theta_-(r - c' t^{\alpha'})) \Psi_t \|^2 = \sprod{\Psi_t}{\Gamma(\theta_-(r - c' t^{\alpha'})) \Psi_t} \to 0
\end{align*} 
for $\Psi = f(H_+) \Phi$ and $\Phi$ in the dense subspace $D$ given by \Cref{min-esc}. From the obvious inequality $\prod\theta_{-}(x_i) \leq \theta_{-}(\sum x_i)$, from $N=d\Gamma(1)$ and from $B=d\Gamma(r)$ we get
\begin{align}  \label{aux: 1}
      \Gamma(\theta_{-}(r-c't^{\alpha'})) \leq \theta_{-}(B-c't^{\alpha'}N).
\end{align} 
We now pick $c \in (c', 1)$ and a smooth function $\rchi$ satisfying $\theta_-(x + (c - c')) \leq \rchi(x)$ as well as the assumptions of \Cref{min-esc}.  It follows that
\begin{align}   \label{aux: 2}
    \theta_{-}(B-c't^{\alpha'}N) \rchi(N\leq t^{1-\alpha'}) \le \theta_{-}(B-c't) = \theta_{-}(B/t-c') \le \rchi(B/t-c).
\end{align} 
Writing $1 = \rchi(N\leq t^{1-\alpha'}) + \rchi(N > t^{1-\alpha'})$ we conclude from \eqref{aux: 1}, \eqref{aux: 2}
$$
    \sprod{\Psi_t}{\Gamma(\theta_{-}(r-c't^{\alpha'})\Psi_t} \le   \sprod{\Psi_t}{\rchi(B/t -c)\Psi_t} +  \sprod{\Psi_t}{\rchi(N > t^{1-\alpha'}) \Psi_t}, 
$$
where the first term, by \Cref{min-esc}, is $O(t^{\nu-1})$, and the second one is $O(t^{\nu - 1 + \alpha'})$ by G\'erard's bound \eqref{Gerard bound N}. By choice of $\alpha'$ we have $\nu - 1 + \alpha' < 0$ and hence the assertion follows. 
\end{proof}


\section{Proof of \Cref{thm:AC}} \label{main thm proof}

Let $\Psi \in \ran \,  \rchi_\Delta(H_+) \chizero$ with $\rchi(N \geq m) e^{-i H t}\Psi \to 0$ as $m, t \to \infty$. 
Our goal is to prove that $\Psi$ is a scattering state in the sense of \eqref{AC reformulated with H}.
Since $\mu > 1/2$ we may choose $\alpha$ such that $\frac{1}{\mu + 3/2} < \alpha < \frac{1 + \mu}{2 + \mu}$, and hence both constraints on $\alpha$ from \Cref{sec:DS} and
\Cref{sec:min-escape} are satisfied. By \Cref{thm-W} the limit
$$
W\Psi = \lim_{t\to\infty} e^{i\tH t} \uGamma(j_t) e^{-iHt} \Psi
$$
exists with $j_t = j_{0, t} \oplus j_{\infty, t}$ satisfying \eqref{def-j}.
Since $\uGamma(j_t)$ is an isometry, it follows that
\begin{align}
    e^{-iHt}\Psi &= \uGamma(j_t)^{*} \uGamma(j_t)e^{-iHt}\Psi \notag\\
       &= \uGamma(j_t)^{*}  e^{-i\tH t}W\Psi + o_t(1). \label{ac1}
\end{align}

Our first goal is to establish \eqref{ac5}, below.
In view of \eqref{pull through},  $\Psi = \rchi_\Delta(H_+) \Psi$, and $\Psi = \rchi_{\{ 0 \}}(H_-) \Psi$ we have
\begin{align}  
W\Psi &= \rchi_\Delta(\tH_{+})W\Psi \label{tH plus energy} \\
W\Psi &= \rchi_{\{ 0 \}}(\tilde{H}_-) W \Psi =\left[ \chizero \otimes  \chizero \right] W\Psi. \label{no neg bosons}
\end{align}
By definition, $\tH_{+} = H_{+} \otimes 1 + 1 \otimes d\Gamma(s_{+})$, where $d\Gamma(s_{+})\ge 0$. It follows that
\begin{equation}\label{ac2}
      \rchi_\Delta(\tH_{+}) = \Big(\rchi_{\Delta}(H_{+})\otimes 1 \Big) \rchi_\Delta(\tH_{+})
\end{equation}
and, for $\Delta' = [E+\eps, \lambda - \eps) \subset \Delta$,
\begin{equation}\label{ac3}
     \rchi_{\Delta}(H_{+}) = \rchi_{\{E\}}(H_{+}) +  \rchi_{\Delta'}(H_{+}) + o_{\eps}(1),
\end{equation}
where $o_{\eps}(1)\to 0$ in the strong operator topology, as $\eps\to 0$. Here the interval $\Delta'$ with $\overline{\Delta'} \subset (E, \lambda)$ is chosen to meet
the hypotheses of \Cref{min-escape} on minimal escape. From \eqref{tH plus energy}, \eqref{ac2} and \eqref{ac3} we see that
$$
    W\Psi = \Big(\rchi_{\{E\}}(H_{+})\otimes 1 \Big) W\Psi + \Big(\rchi_{\Delta'}(H_{+})  \otimes 1 \Big) W\Psi + o_\eps(1).
$$
Notice that, by \eqref{no neg bosons}, the vector $W\Psi$ contains no bosons of negative energy.
On the right of the above equation we can therefore approximate $W\Psi$ by a vector $\Psi_\eps\in (\HHel \otimes \DD_+) \otimes \DD_+$, where $\DD_+ \subset \FF_\mathrm{fin}(\h)$ is the linear span of vectors of the form
\begin{align*}  
a^*(h_1) ... a^*(h_n) \Omega, \quad h_1, ..., h_n \in C_0^\infty((0, \infty) \times S^2).
\end{align*}
This gives another $o_{\eps}(1)$-error and shows that \eqref{ac1} becomes
\begin{align}
    e^{-iHt}\Psi =&\ \uGamma(j_t)^{*}  \Big(e^{-i E t} \rchi_{\{E\}}(H_{+})\otimes e^{-i d\Gamma(s) t} \Big)\Psi_\eps \notag\\
    &+  \uGamma(j_t)^{*}  \Big(e^{-i H t} \rchi_{\Delta'}(H_{+}) \otimes e^{-i d\Gamma(s) t} \Big)\Psi_\eps + o_{\eps}(1) + o_t(1),\label{ac5}
\end{align}
where we used that $e^{-i \tH t} = e^{-i H t} \otimes e^{-i d\Gamma(s)t}$ and $e^{-i H t} = e^{- i H_+ t} e^{i H_- t}$. To conclude the proof, it remains to show that, in the limit $t\to\infty$, the second term vanishes, while in the first term, the operator $\uGamma(j_t)^{*}$ may be replaced by the scattering identification $I$.

To deal with the second term of \eqref{ac5} we choose functions $\tilde{j}_{0,t}, \tilde{j}_{\infty,t}: \R \to [0,1]$ such that  $j_{0,t}\tilde{j}_{0,t} = j_{0,t}$, $j_{\infty,t}\tilde{j}_{\infty,t} = j_{\infty,t}$ and $\tilde{j}_{0, t}(r) \leq \rchi(r \leq t^\alpha)$.
We then have, by \Cref{Gamma adjoint} (ii), 
\begin{equation}\label{ac7}
          \uGamma(j_t)^{*} =  \uGamma(j_t)^{*} \Big(\Gamma(\tilde{j}_{0,t})\otimes \Gamma(\tilde{j}_{\infty,t})\Big).  
\end{equation}
From \Cref{min-escape} it follows that $\Gamma(\tilde{j}_{0,t})e^{-iHt} \rchi_{\Delta'}(H_{+}) \to 0$ in the strong sense as $t\to\infty$. 
This implies that 
\begin{equation*}
    \Big(  \Gamma(\tilde{j}_{0,t}) e^{-i H t} \rchi_{\Delta'}(H_{+}) \otimes \Gamma(\tilde{j}_{\infty,t}) e^{-i d\Gamma(s) t} \Big)\Psi_\eps \to 0\qquad (t\to\infty),
\end{equation*}
which, in view of \eqref{ac7}, shows that the second term of \eqref{ac5} is $o_t(1)$.

It remains to show that $\big(  \uGamma(j_t)^{*} - I\big)\Phi_t \to 0$ with 
$$
     \Phi_t \coloneqq   \Big(e^{-i E t}\rchi_{\{E\}}(H_{+})\otimes e^{-i d\Gamma(s) t} \Big)\Psi_\eps.
$$
We first argue that it suffices to prove this with a boson number cutoff in front. Indeed, with $\tilde{N} = N \otimes 1 + 1 \otimes N$ in $\tHH$ it follows that $\tilde{N} \uGamma(j_t) = \uGamma(j_t) N$ and therefore
$$
     \| \rchi(N \ge m)  \uGamma(j_t)^{*} \Phi_t\| = \| \uGamma(j_t)^{*} \rchi(\tilde{N}\ge m) \Phi_t\| \leq \|\rchi(\tilde{N}\ge m) \Phi_{t=0}\| = o_m(1).
$$
By an analog of  \Cref{AC implies N} (i), $\sup_{t} \| \rchi(N \ge m) I \Phi_t\| = o_m(1)$. Hence it suffices to prove that for every $m \in \N$
$$
     \rchi(N < m) \big(  \uGamma(j_t)^{*} - I\big)\Phi_t \to 0 \qquad (t \to \infty).
$$
We use $\uGamma(j_t)^{*} = I(\Gamma(j_{0, t})\otimes \Gamma(j_{\infty, t}))$ and $\|\rchi(N < m) I\| \le 2^{m/2}$, see \Cref{Gamma adjoint}. So
\begin{eqnarray*}
   \lefteqn{ 2^{-m/2}\| \rchi(N < m) \big(  \uGamma(j_t)^{*} - I\big) \Phi_t \| }\\ 
     && \le \|(\Gamma(j_{0, t})\otimes \Gamma(j_{\infty, t}) - 1) \Phi_t\|\\
      && \le  \| (\Gamma(j_{0, t})-1) \otimes \Gamma(j_{\infty, t})\Phi_t\| +  \|1\otimes (1-\Gamma(j_{\infty, t}))\Phi_t \|.
\end{eqnarray*}
The fact that $j_0(r/t^{\alpha}) \to 1$ as $t\to\infty$, for all $r\in \R$, implies that $(\Gamma(j_{0, t})-1) \xrightarrow{\hspace{1.0mm} s \hspace{1.0mm}} 0$. In view of the trivial time dependence of $\Phi_t$, this shows that
$$
       \| (\Gamma(j_{0, t})-1) \otimes \Gamma(j_{\infty, t})\Phi_t\| \to 0.
$$
On the other hand, $j_{\infty, t} e^{-ist} = e^{-ist} j_\infty^{t} $ with $j_\infty^{t}(r):= j_\infty((r+t)/t^{\alpha}) \to 1$ for all $r\in \R$. This implies 
$(1-\Gamma(j_{\infty, t})) e^{-id\Gamma(s)t} \xrightarrow{\hspace{1.0mm} s \hspace{1.0mm}} 0$ and hence
$$
   \|1\otimes (1-\Gamma(j_{\infty, t}))\Phi_t \| \to 0.
$$

In summary, we have shown that 
\begin{align*}
    e^{-i H t} \Psi =& I e^{-i \tH t} \Big(\rchi_{\{E\}}(H_{+}) \chizero \otimes \chizero \Big)\Psi_\eps + o_{\eps}(1) + o_t(1)
\end{align*}
with $o_\eps(1) \to 0$ as $\eps \to 0$ uniformly in $t$ and $o_t(1) \to 0$ as $t \to \infty$ for each $\eps$. Hence \eqref{AC reformulated with H} holds for $\Psi$.

\medskip

\noindent
\emph{Acknowledgement} M.G. thanks J\'er\'emy Faupin for the hospitality at 
Universit\'e de Lorraine, Metz, in June 2025, and for helpful discussions and explanations concerning  \cite{FauSig2014}.
This work is supported by the German Research Foundation (DFG) under Grant GR 3213/4-1.


\appendix
\section{Fock Space and Second Quantization}
\label{sec:Fock}
In this section we collect basic facts on second quantization. For a more elaborate exposition and proofs we refer to \cite{DG1999}. 

\subsection{Basic definitions}

Let $\h$ denote a one-particle Hilbert space. Let $\FF(\h)$ be the boson Fock space over $\h$. We denote with $a^*(h)$ and $a(h)$ the usual creation and annihilation operators in $\FF(\h)$ satisfying the CCR
\begin{align*}  
[a(g),a^{*}(h)] = \sprod{g}{h},\quad
[a(g),a(h)] =0, \quad [a^{*}(g),a^{*}(h)] =0 \quad (g, h \in \h). 
\end{align*} 
Here, and throughout this paper, the inner product is anti-linear in the first and linear in the second argument.
Let 
$$\phi(h) = a(h) + a^*(h) \quad (h \in \hh),$$
which is essentially self-adjoint on the subspace of finite particle vectors in $\FF(\h)$. If $\omega$ is a self-adjoint operator in $\h$ and $h$ is in the domain of $\omega$ then 
\begin{align*}  
	i [d\Gamma(\omega), \phi(h)] = \phi(i \omega h).
\end{align*} 
Let $j_0, j_\infty$ be bounded operators in $\hh$ satisfying $j_0^* j_0 + j_\infty^* j_\infty = 1$. Then the operator
\begin{align}  \label{appendix j}
j = j_0 \oplus j_\infty: \hh \to \hh \oplus \hh,
\quad h \mapsto (j_0 h, j_\infty h),
\end{align} 
satisfies $j^* j = j_0^* j_0 + j_\infty^* j_\infty = 1$. It follows that $\Gamma(j): \FF(\h) \to \FF(\h \oplus \h)$ is a bounded operator with $\Gamma(j)^* \Gamma(j) = 1$. We have
\begin{align}  
	\Gamma(j) \phi(h) &= \phi(j h) \Gamma(j) \qquad (h \in \hh), \label{a1} \\ 
	d\Gamma(\omega \oplus \omega) \Gamma(j) - \Gamma(j) d\Gamma(\omega) &=   d\Gamma(j, [\omega, j]), \label{a2}
\end{align} 
where $[\omega, j] \coloneqq  [\omega, j_0] \oplus [\omega, j_\infty]$. The (possibly unbounded) operator $d\Gamma(j, k): \FF(\h) \to \FF(\h \oplus \h)$ is defined on the $n$-particle sector of $\FF(\h)$ by
\begin{align*}  
	d\Gamma(j, k) = \sum_{i = 1}^n j \otimes ... \otimes j \otimes \underbrace{k}_{i \mathrm{th}} \otimes \,  j \otimes ... \otimes j,
\end{align*} 
and $d\Gamma(j, k) = 0$ on the vacuum sector. 

\subsection{Factorizing the Fock space}

We define the \textit{canonical unitary} 
\begin{align}  \label{canonical unitary}
	U: \FF(\h \oplus \h) \to \FF(\h) \otimes \FF(\h) 
\end{align} 
on the linear span of vectors of the form $a^*(h_1) ... a^*(h_n) \Omega$, $h_1, ..., h_n \in \hh \oplus \hh$, by setting
\begin{align*}  
	U \Omega &= \Omega \otimes \Omega  \\ 
	U a^*(h) &= \big( a^*(h_0) \otimes 1 + 1 \otimes a^*(h_\infty) \big) U \qquad (h = (h_0, h_\infty) \in \hh \oplus \hh). 
\end{align*} 
Here, $\Omega$ denotes the vacuum in Fock space. From the CCR it follows that $U$ is isometric. The closure of $U$ is unitary. Moreover, 
\begin{align}  
U \phi(h) &= \big( \phi(h_0) \otimes 1 + 1 \otimes \phi(h_\infty) \big) U \hspace{4.5mm} \quad (h = (h_0, h_\infty) \in \hh \oplus \hh), \label{a5}\\ 
U d\Gamma(\omega_0 \oplus \omega_\infty) &= \big( d\Gamma(\omega_0) \otimes 1 + 1 \otimes d\Gamma(\omega_\infty) \big) U. \label{a6}
\end{align} 
We set
\begin{align*}  
	\uGamma(j) \coloneqq U \Gamma(j): \FF(\h) \to \FF(\h) \otimes \FF(\h).
\end{align*} 
Then, by \eqref{a1} and \eqref{a5},
\begin{align*}  
	&\uGamma(j) \phi(h) = \big( \phi(j_0 h) \otimes 1 + 1 \otimes \phi(j_\infty h)  \big)  \uGamma(j) \qquad (h \in \hh), 
\end{align*} 
and, by \eqref{a2} and \eqref{a6},
\begin{align*}  
	&\big( d\Gamma(\omega) \otimes 1 + 1 \otimes d\Gamma(\omega) \big) \uGamma(j) - \uGamma(j) d\Gamma(\omega) = d\uGamma(j, [\omega, j]),
\end{align*} 
where the notation $d\uGamma(j, k) \coloneqq U d\Gamma(j, k)$ was introduced. 

\begin{lemma}\label{lm:udGamma}
Let $j$ be the operator \eqref{appendix j}. Let $k = k_0 \oplus k_\infty: \h \to \hh \oplus \hh$ with self-adjoint operators $k_0$ and $k_{\infty}$. Then for all $u \in \FF(\h)\otimes \FF(\h)$ and all $v\in \FF(\h)$
\begin{align*}  
	 |\sprod{u}{\udGamma(j,k)v}| \leq \| &(d\Gamma(|k_0|) \otimes 1)^{1/2} u \| \, \| d\Gamma(|k_0|)^{1/2} v \| \\ 
	 &+ \| (1 \otimes d\Gamma(|k_\infty|))^{1/2} u \| \, \| d\Gamma(|k_\infty|)^{1/2} v \|.
\end{align*} 
\end{lemma}

For the proof see Lemma 2.16 iv) in \cite{DG1999}.

\subsection{The Scattering Identification}
Let $\iota: \hh \oplus \hh \to \hh$ be given by $\iota (h_0, h_\infty) = h_0 + h_\infty$. We define the \textit{scattering identification} 
\begin{align}  \label{scattering identification}
	I \coloneqq \Gamma(\iota) U^*: U D(\Gamma(\iota)) \subset \FF(\hh) \otimes \FF(\hh) \to \FF(\hh).
\end{align} 
The operator $I$ collects the bosons from the two Fock spaces in the sense that 
$$
	I \big( a^*(g_1) ... a^*(g_n) \Omega \otimes a^*(h_1) ... a^*(h_m) \Omega \big) = a^*(g_1) ... a^*(g_n) a^*(h_1) ... a^*(h_m) \Omega.
$$
\begin{lemma}\label{Gamma adjoint} 
\begin{enumerate}
\item[(i)] For every $m \in \N$ we have $\| \rchi(N \leq m) I \| \leq 2^{m/2}$.
\item[(ii)]  Let $j$ be the operator \eqref{appendix j}. Then $\uGamma(j)^{*} = I\big(\Gamma(j_0^{*})\otimes \Gamma(j_{\infty}^{*}) \big)$.
\end{enumerate}
\end{lemma}
\begin{proof}
(i) Since  $\| \iota \| = 2^{1/2}$, it follows that
$$\| \rchi(N \leq m) I \| = \|\rchi(N \leq m) \Gamma(\iota)\| \leq 2^{m/2}.$$
 (ii) From $\uGamma(j) = U \Gamma(j)$ and $j^{*} = \iota \circ (j_0^{*}\oplus j_\infty^{*})$ it follows that
\begin{align*}
    \uGamma(j)^{*} &= \Gamma(\iota)\Gamma(j_0^{*}\oplus j_\infty^{*}) U^*\\
     &= \Gamma(\iota)U^{*}\cdot U \Gamma(j_0^{*}\oplus j_\infty^{*})U^{*} = I\big(\Gamma(j_0^{*})\otimes \Gamma(j_{\infty}^{*}) \big). \qedhere
\end{align*}
\end{proof}


\section{The Helffer-Sjöstrand formula} 
\label{sec:HS formula}
Let $A$ be a self-adjoint operator. For $f \in C_0^\infty(\R)$ the operator $f(A)$, defined by functional calculus, can be expressed in terms of 
\begin{align}  
	f(A) = \int (z - A)^{-1} d \tilde{f}(z),
\label{DEFINITION HS FORMULA}
\end{align} 
where $\tilde{f} \in C_0^\infty(\C)$ is an \textit{almost analytic extension} of $f$ \cite{D1995}. The integral is taken over $z = x + i y \in \C \cong \R^2$ and we use the abbreviation
$$d \tilde{f}(z) = -\frac{1}{2 \pi} \frac{\partial \tilde{f}}{\partial \overline{z}}(z) dx dy, \quad  \frac{\partial \tilde{f}}{\partial \overline{z}}= \frac{\partial \tilde{f}}{\partial x} + i \frac{\partial \tilde{f}}{\partial y}.$$
The extension $\tilde{f}$ satisfies the Cauchy-Riemann equations on the real axis, 
$$\frac{\partial \tilde{f}}{\partial \overline{z}}(z) = 0 \quad \text{for} \: z  \in \R.$$ 
It is important that we may pick $\tilde{f}$ such that $\frac{\partial \tilde{f}}{\partial \overline{z}}$ vanishes sufficiently fast on the real axis \cite{D1995}; for each $n \in \N$ we may pick $\tilde{f}$ such that  
\begin{align}  \label{real axis decay}
	\left| \frac{\partial \tilde{f}}{\partial \overline{z}}(z) \right| \leq C |y|^n.
\end{align} 

The following lemma is needed to make use of \eqref{DEFINITION HS FORMULA} in cases where $f \in C^\infty(\R)$ is not compactly supported.
\begin{lemma}\label{analytic extension}
Let $f \in C^\infty(\R)$ and suppose for all $n \in \N_0$ 
\begin{align}  
	|f^{(n)}(x)| \leq C_n \expect{x}^{-n}.
\label{decay of function f}
\end{align} 
Let $\eta \in C_0^\infty(\R)$ with $\eta = 1$ in a neighborhood of $0$. For $\varepsilon > 0$ let $\eta_\varepsilon(x) = \eta(\varepsilon x)$  and $f_\varepsilon = f \eta_\varepsilon \in C_0^\infty(\R)$. Then for every $n \in \N$ there exists an almost analytic extension $\tilde{f_\varepsilon} \in C_0^\infty(\C)$ of $f_\varepsilon$ such that uniformly in $\varepsilon$
\begin{align*}  
\supp \tilde{f_\varepsilon} &\subset \{ z \in \C | \,  |y| \leq 2 \expect{x} \}, \\
\left| \frac{\partial \tilde{f_\varepsilon}}{\partial \overline{z}}(z) \right| &\leq C  \expect{x}^{-1 - n} |y|^n.
\end{align*} 
\end{lemma}

\begin{proof} The following construction is similar to the one given in Lemma B.2 of \cite{FauSig2014}. Choose $\gamma \in C_0^\infty(\R)$ with $\gamma(y) = 1$ for $|y| < 1$ and $\gamma(y) = 0$ for $|y| > 2$. Then, using \eqref{decay of function f}, it is straightforward to check that 
\begin{align}  
	\tilde{f}(x + i y) = \left( \sum_{k = 0}^n \frac{f^{(k)}(x)}{k!} (i y)^k \right) \gamma(y / \expect{x}) \label{explicit construction of extension}
\end{align} 
is bounded and satisfies 
\begin{align}  
\supp \tilde{f} &\subset \{ z \in \C | |y| \leq 2 \expect{x} \}, \label{HS lemma 1}\\ 
\left| \frac{\partial \tilde{f}}{\partial \overline{z}}(z) \right| &\leq C  \expect{x}^{-1 - n} |y|^n. \label{HS lemma 2}
\end{align} 
Let the extension $\tilde{\eta}$ of $\eta$ be defined by \eqref{explicit construction of extension} as well. Then $\tilde{f_\varepsilon}(z) \coloneqq \tilde{f}(z) \, \tilde{\eta}(\varepsilon z)$ is an almost analytic extension of $f_\varepsilon$. From \eqref{HS lemma 2} applied to $\tilde{f}$ and $\tilde{\eta}$, it follows that 
\begin{align*}  
\left| \frac{\partial \tilde{f_\varepsilon}}{\partial \overline{z}}(z) \right| &\leq | \tilde{f}(z) | \left| \varepsilon \frac{\partial \tilde{\eta}} {\partial \overline{z}}(\varepsilon z) \right| +  \left| \frac{\partial \tilde{f}} {\partial \overline{z}} (z) \right|   | \tilde{\eta}(\varepsilon z) | \\ 
&\leq C\varepsilon \expect{\varepsilon x}^{-1 - n} |\varepsilon y|^n + C \expect{x}^{-1 - n} |y|^n \\ 
&\leq C  \expect{x}^{-1 - n} |y|^n. 
\end{align*} 
\end{proof}

\section{Operator estimates}
\label{sec:operator estimates}
In this section, we collect technical estimates for the operators introduced in \Cref{introduction} and \Cref{sec:expanded}.

\begin{lemma}
\label{field operator estimates} 
\begin{enumerate}
\item[(i)] For $i = 1, ..., n$ let $w_i \in \hh_\mathrm{ph}$ and $\omega^{-1/2} w_i \in \hh_\mathrm{ph}$. Then
\begin{align}  
	\| \phi(w_1) \dots  \phi(w_n) (1 + H_\omega)^{-n/2}\| \leq C_n \|(1 + \omega^{-1/2}) w_1\|\dots \|(1 + \omega^{-1/2}) w_n\|.
\label{phi prod nelson}
\end{align} 
\item[(ii)] For $i = 1, ..., n$ let  $v_i \in \hh$, $v_i = \theta_+ v_i$ with $\theta_+(s) = \rchi(s \geq 0)$ and $s_+^{-1/2} v_i \in \hh$. Then
\begin{align}
	\|\phi(v_1)\dots\phi(v_n) (1 + d\Gamma(s_+))^{-n/2}\| \leq C_n \|(1 + s_+^{-1/2})v_1\|\dots\|(1 + s_+^{-1/2})v_n\|.
\label{phi prod expanded}
\end{align}
\end{enumerate}
\end{lemma}
\begin{proof}
For the proof of (i) see Lemma 17 in \cite{FGSch2001}. Inequality \eqref{phi prod expanded} follows from \eqref{phi prod nelson}: With $w_i\in \hh_\mathrm{ph}$ defined in terms of $v_i$ by $w_i(k) \coloneqq v_i(|k|, \hat{k})/|k|$, the right-hand sides of \eqref{phi prod nelson} and \eqref{phi prod expanded} agree. We claim that the left-hand sides agree as well. Indeed, with the unitary $\WW$ from \Cref{sec:expanded} and \eqref{U1}, \eqref{U2}, \eqref{V1} it follows that 
 \begin{align*}  
 \WW \big( \phi(w_1)\dots\phi(w_n) (1 + H_\omega)^{-n/2} \otimes 1 \big) = \phi(v_1)\dots\phi(v_n) (1 + d\Gamma(s_+))^{-n/2} \WW. 
 \end{align*} 
\end{proof}

\begin{lemma} \label{N f N inverse lemma}
Let $f \in C_0^\infty(\R)$. Then for all $z \in \C \backslash \R$ 
\begin{align}  
	\|N R_+(z) f(H_+) (N + 1)^{-1}\|\leq C \frac{1}{| \mathrm{Im}z |^2},
	\label{N f N inverse}
\end{align} 
where $H_+$ is the operator \eqref{DEFINITION H_+} and $R_+(z) = (z - H_+)^{-1}$. 
\end{lemma} 
\begin{proof}
First, we prove that $[N, f(H_+)]$ is bounded. Indeed, from the HS-fromula \eqref{DEFINITION HS FORMULA} it follows that
\begin{align*}  
	[N, f(H_+)] = \int [N, R_+(z)] d\tilde{f}(z) = - \int R_+(z) i g \phi(i v_x) R_+(z) d\tilde{f}(z),
\end{align*} 
with the extension $\tilde{f}$ of $f$ chosen such that $|\partial_{\overline{z}} \tilde{f}(z)|/|\mathrm{Im}z|^2$ is integrable. It follows that $[N, f(H_+)]$ is bounded because $\| \phi(i v_x) (1 + d\Gamma(s_+))^{-1/2} \| < \infty$, see \Cref{field operator estimates}, and $\|(1 + d\Gamma(s_+))^{1/2} R_+(z)\| = O((1 + |z|)/|\mathrm{Im}z|)$. Next, we prove \eqref{N f N inverse}. We have
\begin{align*}  
	N R_+(z) f(H_+) (N + 1)^{-1} &= [N, R_+(z) f(H_+)] (N + 1)^{-1} + R_+(z) f(H_+) N (N + 1)^{-1}, 
\end{align*} 	
where $\| R_+(z) \| = O(1/|\mathrm{Im}z|)$. Since $[N, f(H_+)]$ is bounded, 
\begin{align*}  
	 [N, R_+(z) f(H_+)]  &= [N, R_+(z)] f(H_+) + R_+(z) [N, f(H_+)] \\ 
	 &= R_+(z) [N, H_+] f(H_+) R_+(z) + O(1/|\mathrm{Im}z|) \\ 
	 &= O(1/|\mathrm{Im}z|^2) + O(1/|\mathrm{Im}z|),
\end{align*} 
where in the last line we used that $ [N, H_+] f(H_+) = -i g \phi(i v_x) f(H_+)$ is bounded. 
\end{proof} 

\begin{lemma}[Invariance of domains] \label{invariance of domains}
Let $B = d\Gamma(r)$ with $r = i \partial_s$ in $\hh$. 
Let $\mu > 0$ and  $f \in C_0^\infty(-\infty, \Sigma)$. Then the operator $f(H_{+})$ leaves the subspaces $D(N)$ and $D(N) \cap D(B)$ invariant.
\end{lemma}

\begin{proof}
Without loss of generality, we assume that $f$ is real-valued. The invariance of $D(N)$ under $f(H_+)$ is  a corollary of \Cref{N f N inverse lemma}. We now prove the invariance of $D(N) \cap D(B)$. Let $\Phi \in D(N) \cap D(B)$. Since $B$ is self-adjoint, $f(H_{+}) \Phi \in D(B)$ is equivalent to proving that there exists $C$ such that for all $\Psi \in D(B)$
\begin{align*}  
|\sprod{f(H_{+}) \Phi}{B \Psi}| \leq C  \|\Psi \|.
\end{align*} 
We have
\begin{align*}  
\sprod{f(H_{+}) \Phi}{B \Psi} = \sprod{\Phi}{[f(H_{+}), B] \Psi} + \sprod{B \Phi}{f(H_{+}) \Psi},
\end{align*} 
where the commutator is understood in form sense. Since  $|\sprod{B \Phi}{f(H_{+}) \Psi}| \leq \|B\Phi\| \|f(H_{+})\| \|\Psi\|$, it remains to prove that
\begin{align} \label{inv eq 0}
	|\sprod{\Phi}{[f(H_{+}), B] \Psi}| \leq C \| \Psi \|
\end{align}  
with $C$ independent of $\Psi$. We pick a real-valued $g \in C_0^\infty(-\infty, \Sigma)$ such that $f(H_{+}) = g(H_{+}) f(H_{+})$. Then
\begin{align}  \label{inv eq 2}
	\sprod{\Phi}{[f(H_{+}), B] \Psi} = \sprod{g(H_{+}) \Phi}{ [f(H_{+}), B]\Psi} + \sprod{\Phi}{[g(H_{+}), B] f(H_{+}) \Psi}.
\end{align} 
From the HS-formula \eqref{DEFINITION HS FORMULA} for $h \in \{f, g\}$ it follows that 
\begin{align}  \label{inv eq 1}
	i[h(H_+), B] &= \int R_+(z) i[H_+, B]R_+(z) d\tilde{h}(z) \nonumber \\ 
	&= \int R_+(z) \big( N_+ + g \phi(i r v_x) \big)R_+(z) d\tilde{h}(z),
\end{align} 
where $N_+ = d\Gamma(\theta_+(s))$, and the extension $\tilde{h}$ of $h$ is chosen such that $|\partial_{\overline{z}} \tilde{h}(z)|/|\mathrm{Im}z|^3$ is integrable. 
The $N_+$-term in \eqref{inv eq 1} is estimated in the sense \eqref{inv eq 0} using $N_+ \leq N$,  \Cref{N f N inverse lemma} and $\Phi \in D(N)$. The $\phi(i r v_x)$-terms are estimated using that $\expect{x}f(H_+)$ and $\expect{x}g(H_+)$ are bounded,  combined with \Cref{field operator estimates} (ii) in the form
\begin{align}  \label{inv eq 3}
\|\expect{x}^{-1} \phi(i r v_x) (d\Gamma(s_+) + 1)^{-1} \| \leq C \sup_{x \in \R^3} \expect{x}^{-1}\| (1 + s_+^{-1/2}) r v_x \|.
\end{align}
The right-hand side of \eqref{inv eq 3} is finite because $\mu > 0$ implies $(1 + s_+^{-1/2}) v'(s) \in L^2(\R)$.
\end{proof}

\begin{proposition}[Cauchy criterion] \label{abstract cauchy}
Suppose $H$ and $\tH$ are self-adjoint in $\HH$ and $\tHH$, respectively. Let $\phi(t) \in \LL(\HH, \tHH)$ and suppose that for all $\Psi \in \HH$, $\Phi \in \tHH$, 
\begin{align*}  
\left| \frac{d}{dt} \sprod{\Phi_t}{\phi(t) \Psi_t} \right| \leq \sum_{i = 1}^n \| \tilde{B_i}(t) \Phi_t \| \, \| B_i(t) \Psi_t \| + |\gamma(t)| \| \Phi \| \| \Psi \|,
\end{align*} 
where $\Psi_t = e^{- i H t} \Psi$, $\Phi_t = e^{-i \tH t} \Phi$. If 
\begin{align*}  
	\int_1^\infty \| B_i(t) \Psi_t \|^2 dt &\leq C \| \Psi \|^2, \\ 
	\int_1^\infty \| \tilde{B}_i(t) \Phi_t \|^2 dt &\leq C \| \Phi \|^2, \\ 
	\int_1^\infty |\gamma(t)| dt < \infty,
\end{align*} 
then $\displaystyle s-\lim_{t \to \infty} e^{i \tH t} \phi(t) e^{-i H t}$ exists.
\end{proposition}

\begin{proof}
The assumptions imply that $t \mapsto  \sprod{\Phi}{e^{i \tH t} \phi(t) e^{-i H t} \Psi}$ satisfies the  Cauchy-condition uniformly in $\| \Phi \| = 1$. \qedhere
\end{proof}


\section{Fourier estimates}
\label{sec:aux}
This section contains technical results necessary for estimating expressions that contain operators in both position and momentum space. 

\begin{lemma} \label{short-range lemma} Let $v(s) = s^{\mu + 1} \zeta(s) \theta_+(s)$ with $\mu \geq -1$, $\zeta \in \mathcal{S}(\R)$ a Schwartz function and $\theta_+(s) = \rchi(s \geq 0)$ the Heaviside function. Then 
\begin{align*}  
	\check{v}(r) = \frac{1}{\sqrt{2 \pi}} \int_{-\infty}^\infty v(s) e^{i r s}ds 
\end{align*} 
satisfies
$
	\check{v}(r) = O(r^{- ( \mu + 2)}) 
$ as $r \to \infty$. 
\end{lemma}
\begin{proof}
Write $\mu + 1 = n + \theta$ with $n = \lfloor \mu  \rfloor + 1 \in \N_0$ and $\theta \in [0, 1)$. Integrating by parts $n$ times we find that
\begin{align*}  
	\check{v}(r) = \frac{1}{\sqrt{2 \pi}} \int_0^\infty s^{n + \theta} \zeta(s) e^{i r s} ds = \int_0^\infty s^\theta \gamma(s) e^{i r s} ds \frac{1}{(i r)^n},
\end{align*} 
with a Schwartz function $\gamma \in \mathcal{S}(\R)$. It remains to show that 
\begin{align}  \label{FT lm 1}
 \int_0^\infty s^\theta \gamma(s) e^{i r s} ds = O(r^{-(1 + \theta)}).
 \end{align} 
By explicit computation,
  \begin{align}  \label{FT lm 2}
 	\int_0^\infty s^\theta e^{-s} e^{i r s} ds &= O(r^{-(1 + \theta)}).
 \end{align} 
We define the auxiliary function $\eta(s) \coloneqq \gamma(s) - \gamma(0) e^{-s}$. 
Integrating twice by parts if $\theta > 0$ we find
\begin{align}  
	\int_0^\infty s^{\theta} \eta(s) e^{i r s} ds &=  -\int_0^\infty \left( \theta s^{\theta - 1} \eta(s) + s^{\theta} \eta'(s) \right) e^{i r s} ds \frac{1}{i r} \nonumber \\ 
	&= \int_0^\infty \left( \theta (\theta - 1) s^{\theta - 2} \eta(s) + 2 \theta s^{\theta - 1} \eta'(s) + s^\theta \eta''(s) \right) e^{i r s} ds \frac{1}{(i r)^2} \nonumber \\ 
	&= O(r^{-2}), \label{FT lm 3}
\end{align} 
where we used $\eta(s) = O(s)$ as $s \to 0$ and $\theta > 0$. If $\theta = 0$ then one integration by parts is sufficient. Equations \eqref{FT lm 2} and \eqref{FT lm 3} imply \eqref{FT lm 1}. 
\end{proof}

\begin{lemma}\label{sqrt-Lap} \, \\ 
\vspace{-8mm}
\begin{enumerate}
\item[(i)]
Suppose $j:\R\to\R$ is a bounded, differentiable function with $j'\in C_0^{\infty}(\R)$. Let $f:\R\to\R$ be possibly unbounded and Lipschitz continuous with constant $L$. Let $f(s)$ denote multiplication with $f$ in $L^2(\R, ds)$ and let $j(r)$ be defined with $r = i \frac{d}{ds}$ in $L^2(\R, ds)$. Then $[f(s),j(r)]$ is a bounded operator and
$$
    \|[f(s),j(r)]\| \le \frac{L}{\sqrt{2\pi}} \int |\widehat{j'}(k)|\, dk.
$$ 
\item[(ii)] If $\alpha > 0$ and $j_t(r) = j(r/t^\alpha)$ then
$$
	 \|[f(s),j_t(r)]\| = O(t^{-\alpha}) \qquad (t \to \infty).
$$
\end{enumerate}
\end{lemma}

\begin{proof}
(i) We first consider the case where $j\in C_0^{\infty}(\R)$. Then 
$$
       j(r) = \frac{1}{\sqrt{2\pi}} \int e^{ikr}\hat{j}(k)\, dk,
$$
where the operator $e^{i k r}$ shifts functions in $L^2(\R,ds)$ by $k$. It follows that $[f(s),e^{ikr}] = (f(s)-f(s-k))e^{ikr}$. Hence the assertion follows from 
$|f(s)-f(s-k)|\le L |k|$ and from $|\hat{j}(k)||k| = |\widehat{j'}(k)|$. 

In the case $j'\in C_0^{\infty}(\R)$, where $j$ may have unbounded support, we make the following approximation argument: let $\rchi\in C_0^{\infty}(\R; [0,1])$ with $\rchi(r)=1$ for $|r|\leq 1$ and $\rchi(r)=0$ for $|r|\ge 2$. Let $\rchi_n(r):=\rchi(r/n)$ and $j_n:= j \chi_n$. Then $j_n\in C_0^{\infty}(\R)$ and hence
$$
    \|[f(s),j_n(r)]\| \le \frac{L}{\sqrt{2\pi}} \int |\widehat{j'_n}(k)|\, dk.
$$
For $\ph,\psi\in D(f(s))$ with $\|\ph\|=\|\psi \|=1$ it follows that  
\begin{align} \label{sqrt-Lap 1}
\big | \sprod{f(s)\ph}{j(r) \psi} - \sprod{j(r)\ph}{f(s)\psi}\big |
   &= \lim_{n\to\infty} \big | \sprod{f(s)\ph}{j_n(r)\psi} - \sprod{j_n(r)\ph}{f(s)\psi}\big | \nonumber \\
   &\le \frac{L}{\sqrt{2\pi}} \limsup_{n \to \infty} \int |\widehat{j'_n}(k)|\, dk.
\end{align}
Since $j'$ has compact support, $j(r)$ becomes constant for $\pm r$ large. Therefore, for $n$ sufficiently large,
\begin{equation}\label{jchi-n}
     (j\chi_n)' = j'\chi_n + j\chi_n' = j' + \theta_{+} \chi_n' + \theta_{-} \chi_n',
\end{equation}
where $\theta_{+}$ and $\theta_{-}$ are constant multiples of the characteristic functions of $\R_{+}$ and $\R_{-}$, respectively. From \eqref{jchi-n} it is easy to see that 
\begin{align}  \label{sqrt-Lap 2}
     \int |\widehat{j'_n}(k)|\, dk = \int |\widehat{j'}(k)|\, dk + O(1/n) \qquad (n\to\infty).
\end{align}
From \eqref{sqrt-Lap 1} and \eqref{sqrt-Lap 2} the assertion follows. 

(ii) follows from (i) and $\widehat{j_t'}(k) = \widehat{j'}(t^\alpha k)$. 
\end{proof}


\end{document}